\documentclass[12pt, journal,a4paper,onecolumn]{IEEEtran}
\usepackage{amsmath}
\usepackage{amssymb}
\usepackage{mydefins}
\usepackage[dvips]{graphicx}

\ifCLASSINFOpdf
\else
\fi
\begin{document}

\sloppy

\title{First- and Second-Order Coding Theorems for Mixed Memoryless Channels with General Mixture}

{
\author{Hideki~Yagi, Te~Sun~Han,~and~Ryo~Nomura
\thanks{H.\ Yagi is with the Dept.\ of Communication Engineering and Informatics, The University of Electro-Communications, Tokyo, Japan (email: h.yagi@uec.ac.jp).}
\thanks{T.\ S.\ Han is with the National Institute of Information and Communications Technology (NICT), Tokyo, Japan (email: han@is.uec.ac.jp).}
\thanks{R.\ Nomura is with School of Network and Information, Senshu University, Kanagawa, Japan (email: nomu@isc.senshu-u.ac.jp).}%
}}
\def\bf#1{\boldsymbol{#1}}
\def\vect#1{\boldsymbol{#1}}
\def\QED{\hfill$\Box$}
\definecolor{Bgreen}{rgb}{ .0, .55, .0 }
\definecolor{Red}{rgb}{ 1.0, .0, .0 }
\definecolor{Navy}{rgb}{ 0.0, .0, 1.0 }

\newcommand{\Tr}{\textrm{Tr}}
\newcommand{\E}{\mathsf{E}}	
\newcommand{\V}{\mathsf{V}}	
\newcommand{\ep}{\textrm{ep}}
\newcommand{\cl}{\textrm{cl}}
\newcommand{\Cov}{\textrm{Cov}}		
\newcommand{\Qinv}{Q_\textrm{inv}}	
\newcommand{\Ginv}{G_\textrm{inv}}
\newcommand{\Code}{\mathcal{C}_n}
\newcommand{\PX}{\mathcal{P}(\mathcal{X})}
\newcommand{\PY}{\mathcal{P}(\mathcal{Y})}
\newcommand{\PXY}{\mathcal{P}(\mathcal{X} \rightarrow \mathcal{Y})}
\newcommand{\bartheta}{{\overline{\theta}}}				
\def\bf#1{\boldsymbol{#1}}
\def\vect#1{\boldsymbol{#1}}
\newcommand{\nm}{{m}}
\def\QED{\hfill$\Box$}




\renewcommand{\baselinestretch}{1.00}
\maketitle


\begin{abstract}
This paper investigates the first- and second-order maximum achievable rates of codes with/without cost constraints for mixed {channels} whose channel law is characterized by a general mixture of (at most) uncountably many stationary and memoryless discrete channels.
These channels are referred to as {mixed memoryless channels with general mixture} and
include the class of mixed memoryless channels of finitely or countably memoryless channels as a special case.
For {mixed memoryless channels with general mixture}, the first-order coding theorem which gives a formula for the $\varepsilon$-capacity is established, and then a direct part of the second-order coding theorem is  provided.
A subclass of {mixed memoryless channels} whose component channels can be ordered according to their capacity is introduced, and the first- and second-order coding theorems are established.
It is shown that the established formulas reduce to several known formulas for restricted scenarios. 
\end{abstract}

\IEEEpeerreviewmaketitle

\section{Introduction}

Investigation of the maximum achievable rate of codes whose probability of decoding error does not exceed $\varepsilon \in [0,1)$ for various coding systems has been one of major research topics in information theory.
The first-order optimum rate for channel codes with such a property  is referred to as the $\varepsilon$-\emph{capacity}.  
Inspired by the recent results of second-order coding theorems given, for example, by Hayashi \cite{Hayashi2009} and Polyanskiy, Poor, and Verd\'u \cite{PPV2010} for stationary memoryless channels, this research topic has become of greater importance from both theoretical and practical viewpoints.

It is well-known that stationary memoryless channels with finite input and/or output alphabets have the so-called \emph{strong converse property}, and 
the $\varepsilon$-capacity coincides with the channel capacity ($\varepsilon$-capacity
with $\varepsilon$ = 0) \cite{Wolfowitz78}. On the other hand, allowing a decoding
error probability up to $\varepsilon$, the maximum achievable rate may
be improved for non-stationary and/or non-ergodic channels.
The simplest example is a class of \emph{mixed channels} \cite{Han2003}, also referred to
as averaged channels \cite{Ahlswede68,Kieffer2007} or decomposable channels \cite{Winkelbauer71a},
whose probability distribution is characterized by a mixture
of multiple stationary memoryless channels. This channel is
stationary but non-ergodic and is of theoretical importance when extensions of coding
theorems for ergodic channels are addressed.

For general channels including mixed channels, a general
formula {for} the $\varepsilon$-capacity has been given by Verd\'u and Han [14].
This formula, however, involves limit operations with respect
to code length $n$, and thus is infeasible to compute in
general. 
On the other hand, for mixed channels of uncountably many stationary and memoryless discrete channels, which will be called {general \emph{mixed memoryless channels}}, a single-letter characterization of the channel capacity has been given by Ahlswede \cite{Ahlswede68} for the case without cost constraints and by Han \cite{Han2003} for the case with cost constraints. 
 These characterizations are of importance because the channel capacity may be computed
with complexity independent of $n$.
Recently, Yagi and Nomura \cite{Yagi-Nomura2014a} has provided a single-letter characterization of the
$\varepsilon$-capacity with/without cost constraints for {mixed channels} of at most countably many stationary memoryless channels.
Regarding the $\varepsilon$-capacity for {mixed memoryless channels with general mixture}, however, no characterizations have been given in the literature. 
The regular decomposable channel which consists of memoryless
channels \cite{Winkelbauer71a}, is one of a few examples for which a single-letter characterization of the $\varepsilon$-capacity is known.
In addition, the second-order optimum rate has been {characterized} only for a few classes of mixed memoryless channels such as the mixed channel of two memoryless additive channels \cite{PPV2011a}, the  mixed channel of finitely many stationary and memoryless discrete channels which can be ordered according to their capacities {\cite{Yagi-Nomura2014b}}, and block fading {channels} characterized as {the} mixed channel {consisting of} additive Gaussian noise channels \cite{YDKP2013}.

This paper first gives a single-letter characterization of the $\varepsilon$-capacity with/without cost constraints for {mixed memoryless channels with general mixture} {(Theorem \ref{theo:general_mixed_DMC_e-cap})}. 
The established formula reduces to the one for the channel capacity 
given by \cite{Ahlswede68} and \cite{Han2003} when $\varepsilon$ is zero. 
{The achievability and converse proofs of Theorem \ref{theo:general_mixed_DMC_e-cap} proceed in a parallel manner: (i)  the upper or lower bound on the error probability is characterized by {the} type (empirical distribution) of codewords and (ii) the convergence of a {subsequence} of types to a certain probability distribution is discussed.
Next, a direct coding theorem (achievability) is given for the second-order optimum rate {(Theorem \ref{theo:general_mixed_DMC_achievable_2nd})}.
In the proof of Theorem \ref{theo:general_mixed_DMC_achievable_2nd}, an upper bound on the error probability is derived based on the random coding argument of a fixed type, and it is a key to specify the type of codewords so that the speed of the convergence of {the} mutual information computed by {this type} to the target first-order coding rate is fast enough (cf. Equation \eqref{eq:good_conv_speed}).
For a fixed code, on the other hand,  we cannot guarantee that the speed of the convergence of such mutual information to the target first-order coding rate is fast enough, and this fact has prevented us from establishing {the} converse part of the second-order coding theorem. }
In order to circumvent this problem, we will introduce a subclass of {mixed memoryless channels with general mixture}, called \emph{well-ordered mixed memoryless channels}, {whose} component channels can be ordered as discussed in \cite{Yagi-Nomura2014b}. 
For this channel class, the first- and second-order coding theorems are established.
It is shown that the established formulas reduce to several known formulas for restricted scenarios. 
All coding theorems are proved based on {the} \emph{information spectrum methods} (c.f. \cite{Han2003,Verdu-Han94}). 
In particular, we {use a proof technique} for the converse part such that the proof proceeds based on an arbitrarily chosen converging subsequence of types of codewords, which may simplify even the proof of the second-order coding theorem for stationary memoryless channels such as in \cite{Hayashi2009}.

This paper is organized as follows: The problem addressed in this paper is stated in Sect.\ \ref{sect:problem_formulation}.
We next establish the first-order coding theorem in Sect.\ \ref{sect:general_mixed_DMC_e-cap} and a direct part of the second-order coding theorem in Sect.\ \ref{sect:general_mixed_DMC_achievable_2nd} for {mixed memoryless channels with general mixture}.
These theorems are proved in Sect.\ \ref{sect:proof}; several lemmas used to prove the theorems are first provided in Sect.\ \ref{sect:lemma}, and then proofs of the coding theorems are given in Sect.\ \ref{sect:proof_general_mixed_DMC_e-cap} and \ref{sect:proof_general_mixed_DMC_achievable_2nd}, respectively.
Section \ref{sect:well-ordered_channel} discusses well-ordered mixed memoryless channels, introduced in Sect.\ \ref{sect:introduction_WO_channel}, and the first- and second-order coding theorems are stated in Sect.\ \ref{sect:coding_theorem_WO_channel} along with the proofs in Sect.\ \ref{sect:proof_WO_mixed_DMC_e-cap} and \ref{sect:proof_WO_mixed_DMC_2nd}. 
Some concluding remarks are given in Sect. \ref{sect:conclusion}.

\section{Problem Formulation} \label{sect:problem_formulation}

\subsection{Mixed Memoryless Channel under General {Mixture}}

Consider a channel $W^n: \mathcal{X}^n \rightarrow \mathcal{Y}^n$, {without any assumption on the memory structure,} which stochastically  maps an input sequence ${\vect{x}} \in \mathcal{X}^n$ of length $n$ into an output sequence ${\vect{y}} \in \mathcal{Y}^n$ {of length $n$}. 
Here, $\mathcal{X}$ and $\mathcal{Y}$ denote {\emph{finite} input and output alphabets}, respectively. 
A sequence $\vect{W} := \{ W^n\}_{n=1}^\infty$ of channels $W^n$ is referred to as a \emph{general channel} \cite{Han2003}.

We consider a \emph{mixed channel}\footnote{Mixed channels are also referred to as \emph{averaged channels} \cite{Kieffer2007} or \emph{decomposable channels} \cite{Winkelbauer71a}.} with a general  probability measure \cite[Sect 3.3]{Han2003}.
Let $\Theta$ be an arbitrary {probability space} and assign a general channel $\vect{W}_\theta = \{ W_\theta^n\}_{n=1}^\infty$ to each $\theta \in \Theta$, which are called \emph{component channels} {or simply \emph{components}}.
Here, we assume that each $\vect{W}_\theta$ has the same input alphabet $\mathcal{X}$ and output alphabet $\mathcal{Y}$.
With an arbitrary probability measure $w$ on $\Theta$, we {define} a \emph{mixed channel} $\vect{W} = \{W^n \}_{n=1}^\infty$ with the conditional probability {distribution} given by
\begin{align}
&W^n({\vect{y}}| \vect{x}) = \int_{\Theta} W_\theta^n({\vect{y}}| \vect{x}) dw(\theta) \nonumber \\
&~~~~~~~~~~~~~~~~~~~(\forall n = 1,2,\cdots ; \forall \vect{x} \in \mathcal{X}^n, {\forall \vect{y} \in \mathcal{Y}^n}). \label{eq:mixed_channel_general_measure}
\end{align}

In this paper, we focus on the case where the component channels are stationary {memoryless discrete} channels.
{Then}, a component channel {can be} denoted {simply} by $\vect{W}_\theta = \{ W_\theta {\, : \mathcal{X} \rightarrow \mathcal{Y}} \}$.
A mixed channel given by \eqref{eq:mixed_channel_general_measure} {with} stationary memoryless {discrete} channels {$\vect{W}_\theta = \{ W_\theta  \}$} is referred to as a general \emph{mixed memoryless channel} for simplicity.

Let $\Code$ be a code of length $n$ and the number of codewords $|\Code| = M_n$.
We denote the codeword corresponding  to message $ {i \in \{ 1,2,\ldots, M_n\}}$ by $\vect{u}_i$, i.e., $\Code = \{ \vect{u}_1, \vect{u}_2, \ldots, \vect{u}_{M_n} \}$.
We assume that the decoding region $D_i$ of $\vect{u}_i$ satisfies
\begin{align}
\bigcup_{i = 1}^{M_n} D_i = \mathcal{Y}^n~~\mathrm{and}~~D_i \cap  D_j = \emptyset~~(i \neq j). \label{eq:decoding_region}
\end{align}
%
The \emph{average}  probability of decoding error over ${\vect{W}}$ is defined as
\begin{eqnarray}
\varepsilon_n := \frac{1}{M_n}\sum_{i=1}^{M_n} W^n(D_i^c|\vect{u}_i), \label{eq:ave_prob}
\end{eqnarray}
where $D_i^c$ denotes the complement set of $D_i$ in $\mathcal{Y}^n$.
{Such a} code $\Code$ is referred to as an $(n,M_n, \varepsilon_n )$ code.


We consider a cost function $c_n(\cdot)$ for $\vect{x} = (x_1,x_2, \ldots,x_n) \in \mathcal{X}^n$, defined as
\begin{align}
c_n (\vect{x}) := \frac{1}{n} \sum_{i=1}^n c(x_i),
\end{align}
where $ c : \mathcal{X} \rightarrow [0,\infty)$. 
A sequence $\vect{x}$ is said to satisfy cost constraint $\Gamma$ if
\begin{align}
c_n (\vect{x}) \le \Gamma, \label{eq:cost_constraint_Gamma}
\end{align}
and an $(n, M_n, \varepsilon_n)$ code $\Code$ is said to satisfy cost constraint $\Gamma$ if every codeword $\vect{u}_i \in \Code$ satisfies cost constraint $\Gamma$. 

\smallskip
\begin{e_rema} \label{rema:no_cost_constraint}
If $\Gamma \ge \max_{x \in \mathcal{X}}c(x)$, then \eqref{eq:cost_constraint_Gamma} holds for any $\vect{x} \in \mathcal{X}^n$.
This case corresponds to the coding system without cost constraints, {which is indicated simply by $\Gamma = + \infty$}.
\QED
\end{e_rema}

\subsection{Optimum Coding Rates}

\begin{e_defin}\label{def:1st_achievable}
{\rm
A {first-order} coding rate $R \ge 0$ is said to be {$(\varepsilon|\Gamma)$}-\emph{achievable} if there exists a sequence of $(n,M_n, \varepsilon_n )$ codes {satisfying} cost constraint $\Gamma$
 such that 
\begin{align}
\limsup_{n \rightarrow \infty} {\varepsilon_n} \le \varepsilon ~~~\mathrm{and}~~~\liminf_{n \rightarrow \infty} \frac{1}{n} \log M_n \ge R. \label{eq:1st_order_achievable} 
\end{align}
The supremum of {all} {$(\varepsilon|\Gamma)$-achievable} rates is called the {first-order} {$(\varepsilon|\Gamma)$}-\emph{capacity} and is denoted by ${C_\varepsilon(\Gamma)}$.
{We {also} write as $C_\varepsilon = C_\varepsilon (+ \infty)$ for simplicity.} 
}\QED
\end{e_defin}

{Set $\Gamma_0:= \min_{x \in \mathcal{X}} c(x)$. If $\Gamma < \Gamma_0$, then obviously $C_\varepsilon (\Gamma) = 0$ because no sequences $\vect{x} \in \mathcal{X}^n$ satisfy cost constraint $\Gamma$, and hence no $R > 0$ is $(\varepsilon|\Gamma)$-achievable.}


\smallskip
{Let $M_{n,\varepsilon}^*$ denote the maximum size of codes of length $n$ and error probability less than or equal to $\varepsilon$ satisfying cost constraint $\Gamma$.
The first-order $(\varepsilon | \Gamma)$-capacity indicates that $M_{n,\varepsilon}^*$ behaves as
\[
\log M_{n,\varepsilon}^* = n C_\varepsilon(\Gamma) + o(n)
\] 
for sufficiently large $n$.
For coding systems whose first-order capacity had been characterized, our next target may be to characterize the second-oder term of $\log M_{n,\varepsilon}^*$.
This motivates us to introduce {the} \emph{second-order coding rates}, and its maximum value denoted by $D_\varepsilon(R|\Gamma)$ with respect to the first-order coding rate $R = C_\varepsilon(\Gamma)$ roughly satisfies the relation 
\[
\log M_{n,\varepsilon}^* \simeq  n C_\varepsilon(\Gamma) + \sqrt{n} D_\varepsilon (R|\Gamma) + o\left(\sqrt{n} \right)
 \]
 for sufficiently large $n$.}
Second-order achievable rates and their optimum value are now {formally} defined {as follows}.
\begin{e_defin}\label{def:2nd_achievable}
{\rm
A {second-order} coding rate ${S}$ is said to be ${(\varepsilon, R|\Gamma)}$-\emph{achievable} if there exists a sequence of $(n,M_n, \varepsilon_n )$ codes {satisfying} cost constraint $\Gamma$ such that
\begin{align}
\limsup_{n \rightarrow \infty} {\varepsilon_n} \le \varepsilon ~~~\mathrm{and}~~~\liminf_{n \rightarrow \infty} \frac{1}{\sqrt{n}} \log \frac{M_n}{e^{n {R}}}  \ge {S}. \label{eq:2nd_order_achievable}  
\end{align}
The supremum of {all} {$(\varepsilon, {R}| \Gamma)$}-achievable rates is called the second-order $(\varepsilon, {R} | \Gamma)$-\emph{capacity} and is denoted by ${D_\varepsilon(R|\Gamma)}$.
{We {also} write as $D_\varepsilon(R) = D_\varepsilon (R|+\infty)$ for simplicity.} 
}\QED
\end{e_defin}


\smallskip
\begin{e_rema} \label{rema:our_interest}
It is easily verified that if $R < {C_\varepsilon(\Gamma)}$ then ${D_\varepsilon(R|\Gamma)} = + \infty$ for all $\varepsilon \in [0, 1)$ {from the definition of capacities}. 
Also, if $R > {C_\varepsilon(\Gamma)}$ then ${D_\varepsilon(R|\Gamma)} = - \infty$ for all $\varepsilon \in [0, 1)$. Therefore, only the case $R =  {C_\varepsilon(\Gamma)}$  is of our {main} interest.
\QED
\end{e_rema}

\section{Coding Theorems for General Mixed Memoryless Channel}

\subsection{First-Order Coding Theorems} \label{sect:general_mixed_DMC_e-cap}

The following theorem gives a single-letter characterization {for} the first-order $(\varepsilon|\Gamma)$-capacity {of} {mixed memoryless channels with general mixture}.

\smallskip
\begin{e_theo} \label{theo:general_mixed_DMC_e-cap}
Let $\vect{W}$ be a {general} mixed memoryless channel with {measure} $w$.
For any fixed $\varepsilon \in [0,1)$ {and $\Gamma \ge {\Gamma_0}$}, the {first-order} {$(\varepsilon|\Gamma)$}-capacity is given by
\begin{align}
{C_\varepsilon(\Gamma)} =  \sup_{ {P} : {\E c({X_P}) \le \Gamma}} \sup \left\{ R \, \Big|  \int_{\{ \theta|\, I ({P}, W_\theta) < R \}} dw(\theta) \le \varepsilon  \right\}, \label{eq:e-Cap_general_mixture}
\end{align}
where
{{$X_P$ indicates the input random variable} subject to distribution $P$ on $\mathcal{X}$,} and  $I ({P}, W_\theta)$ denotes the mutual information {with {input} ${P}$ and channel $W_\theta: \mathcal{X} \rightarrow \mathcal{Y}$} (cf.\ Csisz{\'a}r and K{\"{o}}rner \cite{Csiszar-Korner2011}).
\QED
\end{e_theo}
The proof of this theorem is given in Sect.\ \ref{sect:proof}.

\smallskip
\begin{e_rema}
If $\Theta$ is a singleton, Theorem \ref{theo:general_mixed_DMC_e-cap} reduces to the well-known formula 
\begin{align}
C_\varepsilon(\Gamma) = \sup_{ {P} : {\E c({X_P})} \le \Gamma}  I(P,W)~~(0 \le \forall \varepsilon < 1),
\end{align}
{which means that} the strong converse holds in this case (cf.\ \cite{Csiszar-Korner2011,Wolfowitz78}), unlike in the general case $|\Theta|> 1$. 
{For $\Theta$ which is a finite or countable infinite set, formula \eqref{eq:e-Cap_general_mixture} of the first-order capacity $C_\varepsilon(\Gamma)$  reduces to the formula given by Yagi and Nomura \cite{Yagi-Nomura2014a}.} 
{For {mixed memoryless channels with general mixture}, on the other hand, in the special case {of} $\varepsilon= 0$, formula \eqref{eq:e-Cap_general_mixture} reduces to 
\begin{align}
{C_0(\Gamma)} =  \sup_{ {P} : {\E c({X_P}) \le \Gamma}} w{\rm \mathchar`-ess.inf} I ({P}, W_\theta), \label{eq:0-Cap_general_mixture}
\end{align}
which coincides with the formula given by Han \cite[Theorem 3.6.5]{Han2003}, 
where $w{\rm \mathchar`-ess.inf}$ denotes the essential infimum of $I ({P}, W_\theta)$ with respect to the probability measure $w$.
}
\QED
\end{e_rema}

When $\Theta$ is a singleton, it is known that the $C_{\varepsilon}(\Gamma)$ is concave in $\Gamma$ and is strictly increasing over the range $\Gamma_0 \le \Gamma \le \Gamma^*$, where $\Gamma^*$ denotes the smallest $\Gamma$ at which  $C_{\varepsilon}(\Gamma)$ coincides with $C_{\varepsilon}$ (without cost constraints) (cf.\ Blahut \cite{Blahut72}).
For the case of $|\Theta| > 1$, $C_{\varepsilon}(\Gamma)$ is indeed non-decreasing, but there are examples of mixed memoryless channels for which $C_{\varepsilon}(\Gamma)$ is not strictly increasing in $\Gamma_0 \le \Gamma \le \Gamma^*$.
This also indicates that $C_{\varepsilon}(\Gamma)$ need not be concave in $\Gamma$.

\smallskip
{In the case without cost constraints, Theorem \ref{theo:general_mixed_DMC_e-cap} reduces to the following corollary.}
\begin{e_coro} \label{coro:general_mixed_DMC_e-cap}
Let $\vect{W}$ be a {general} mixed memoryless channel with {measure} $w$.
For any fixed $\varepsilon \in [0,1)$, the {first-order} {$\varepsilon$}-capacity is given by
\begin{align}
{C_\varepsilon} =  \sup_{P} \sup \left\{ R \, \Big|  \int_{\{ \theta|\, I ({P}, W_\theta) < R \}} dw(\theta) \le \varepsilon  \right\}, \label{eq:e-Cap_general_mixture_without_constraints}
\end{align}
where {$\displaystyle \sup_{P}$ denotes the supremum over the set $\PX$} of all probability distributions on $\mathcal{X}$.
\QED
\end{e_coro}

{
\begin{e_rema}
The direct part of formula \eqref{eq:e-Cap_general_mixture_without_constraints} was first demonstrated by Han \cite[Lemma 3.3.3]{Han2003}.
In the special case of $\varepsilon = 0$, we have an alternative formula of $C_0$ as in \eqref{eq:0-Cap_general_mixture} (by replacing the supremum over $\{P \, | \, \E c(X_P) \le \Gamma\}$ with the supremum over $\PX$), which coincides with the formula given by Ahlswede  \cite{Ahlswede68}. See also \cite[Remark 3.3.3]{Han2003} for the equivalence between these characterizations.
\QED
\end{e_rema}
}

\subsection{Second-Order Coding Theorems} \label{sect:general_mixed_DMC_achievable_2nd}

We {now} turn to analyzing second-order coding rates. 
Let $\Psi_{\theta, {P}}$ denote the Gaussian cumulative distribution function with zero mean and variance 
\begin{align}
V_{\theta, {P}} := \sum_{x \in \mathcal{X}}  \sum_{y \in \mathcal{Y}} {P}(x) W_\theta(y|x) \left( \log \frac{W_\theta(y|x)}{{P}W_{\theta}(y)} - D(W_\theta(\cdot|x) || {P}W_{\theta}) \right)^2,
\end{align}
{that is, 
\begin{align}
\Psi_{\theta, P} (z) := G \left( \frac{z}{\sqrt{V_{\theta, P}}} \right), ~~
G(z) := \frac{1}{\sqrt{2\pi}} \int_{-\infty}^z e^{-\frac{t^2}{2}} dt, \label{eq:Gaussian_dist}
\end{align}
}
where 
\begin{align}
{P}W_{\theta}(y) := \sum_{x} {P}(x) W_\theta (y|x)
\end{align}
{denotes the output distribution on $\mathcal{Y}$ due to the input distribution $P$ on $\mathcal{X}$ via channel $W_\theta$}, {and $D(W_\theta(\cdot|x) || {P}W_{\theta})$ denotes the divergence between $W_\theta(\cdot|x)$ and $PW_\theta$.}
{It is known that there are stationary memoryless channels $W_\theta$ for which $V_{\theta, P} = 0$ for some $P \in \mathcal{P}(\mathcal{X})$ (cf. \cite{PPV2010,Strassen62}).
In such a case, with an abuse of notation, we interpret $\Psi_{\theta, P}(z) = G(z/\! \sqrt{V_{\theta,P}})$ as the step function which is defined to take zero for $z < 0$ and one otherwise.}

\smallskip
For the second-order coding rate, we have the following direct theorem (achievability).
\begin{e_theo}[Direct Part] \label{theo:general_mixed_DMC_achievable_2nd}
Let $\vect{W}$ be a {general} mixed memoryless channel with {measure} $w$.
For $\varepsilon \in [0,1)$, $\Gamma \ge {\Gamma_0}$, {and $R \ge 0$}, {it holds that}
\begin{align}
{D_\varepsilon(R|\Gamma)}  \ge  \sup_{ P :  \E c({X_P}) \le \Gamma } \sup \left\{ S \, \Big| \, G_w(R, S|{P})  \le  \varepsilon  \right\} =: \overline{D}_\varepsilon (R | \Gamma), \label{eq:2nd_order_general_mixture}
\end{align}
{where}
\begin{align}
G_w(R, {S} |{P}) :=   \int_{\{ \theta|\, I ({P}, W_\theta) < R \}} dw(\theta) + \int_{\{ \theta|\, I ({P}, W_\theta) = R\}} \Psi_{\theta, {P}}(S)dw(\theta).   \label{eq:func_Gw}
\end{align}
\QED
\end{e_theo}
The proof of this theorem is given in Sect.\ \ref{sect:proof}.

\smallskip
\begin{e_rema} \label{rema:canonical_rep}
The two terms on the right-hand side of \eqref{eq:func_Gw} can be summarized into the following single term:
\begin{align}
\int_{\Theta} dw(\theta) \lim_{n \rightarrow \infty} \Psi_{\theta, P} \big( \sqrt{n} (R - I(P, W_\theta)) + S \big),
\end{align}
which is called the \emph{canonical representation} (cf.\ Nomura and Han \cite{Nomura-Han2013,Nomura-Han2014}).
Let us here focus on the crucial case of $R = C_\varepsilon(\Gamma)$.
In view of formula \eqref{eq:e-Cap_general_mixture} for the $\varepsilon$-capacity $C_\varepsilon(\Gamma)$ it is not difficult to check that, for any $P$ such that $\E c(X_P) \le \Gamma$, 
\begin{align}
{\int_{\{ \theta | I(P, W_\theta) < C_\varepsilon (\Gamma)\}} dw(\theta)} \, & \, {\le \varepsilon,} \label{eq:pre_majorization} \\  
\int_{\{ \theta | I(P, W_\theta) \le C_\varepsilon (\Gamma)\}} dw(\theta) &\ge  \varepsilon \label{eq:post_majorization}
\end{align}
hold.
Thus, we may consider  the following canonical equation for $S$:
\begin{align}
\int_{\Theta} dw(\theta) \lim_{n \rightarrow \infty} \Psi_{\theta, P} \big( \sqrt{n} (C_\varepsilon(\Gamma) - I(P, W_\theta)) + S \big) = \varepsilon. \label{eq:canonical_equation0}
\end{align}
Notice here, in view of \eqref{eq:pre_majorization}  and \eqref{eq:post_majorization}, that equation \eqref{eq:canonical_equation0} always has a solution. 
Let $S_P(\varepsilon)$ denote the solution of this equation, {where}
$S_P(\varepsilon) = + \infty$ if the solution is not unique (notice that this case occurs if $\int_{\{ \theta \, |\, I(P, W_\theta) = C_\varepsilon(\Gamma)\}} dw(\theta) = 0$, which equivalently means that the second term on the right-hand side in \eqref{eq:func_Gw} is zero).
Then, the $\overline{D}_\varepsilon \big(C_\varepsilon(\Gamma)| \Gamma\big)$ (i.e., $R = C_\varepsilon(\Gamma)$) in \eqref{eq:2nd_order_general_mixture} can be rewritten in a simpler form as 
\begin{align}
\overline{D}_\varepsilon \big(C_\varepsilon(\Gamma) | \Gamma\big) = \sup_{ P :  \E c({X_P}) \le \Gamma } S_P(\varepsilon).
\end{align}
We sometimes prefer this simple expression rather than in \eqref{eq:2nd_order_general_mixture}.%
\QED
\end{e_rema}

\begin{e_rema}
Denote the right-hand side of \eqref{eq:2nd_order_general_mixture} again by $\overline{D}_\varepsilon(R|\Gamma)$.
If $\Theta$ is a singleton, it can be easily verified that
\begin{align}
{\overline{D}_\varepsilon(R|\Gamma)} = \left\{
\begin{array}{ll}
- \infty & \mathrm{if}~ R >  C_\varepsilon(\Gamma) \\
 \displaystyle { \sup_{\substack{ P: I(P,W) =R \\ \E c({X_P}) \le \Gamma}  }}
\sup \left\{ S \, \Big| \, \Psi_{P} (S) \le  \varepsilon  \right\} & \mathrm{if}~R = C_\varepsilon (\Gamma) \\
+ \infty & \mathrm{if}~ R < C_\varepsilon(\Gamma),
\end{array}
\right.
\end{align}
{where setting the singleton set $\Theta$ as $\Theta = \{ \theta_0\}$ we use $\Psi_P$ instead of $\Psi_{P,\theta_0}$.}
In particular, if
\begin{align}
R=C_\varepsilon(\Gamma) = \sup_{\substack{ P: I(P,W) =R \\ \E c({X_P}) \le \Gamma}  } I(P,W) ,
\end{align}
then it follows from Theorem \ref{theo:WO_mixed_DMC_2nd} with $|\Theta| = 1$ later in Sect.\ \ref{sect:well-ordered_channel} that
\begin{align}
{D_\varepsilon \big(C_\varepsilon(\Gamma)|\Gamma \big)} = {\overline{D}_\varepsilon \big(C_\varepsilon(\Gamma)|\Gamma\big) } = \left\{
\begin{array}{ll}
\sqrt{V_{\max}}  G^{-1}(\varepsilon) & \mathrm{if}~\varepsilon \ge \frac{1}{2} \\
\sqrt{V_{\min}}  G^{-1}(\varepsilon) & \mathrm{if}~\varepsilon < \frac{1}{2},
\end{array}
\right. \label{eq:second_order_formula}
\end{align}
where 
\begin{align}
V_{\max} &:= \max_{\substack{  P: I(P,W)=C_\varepsilon(\Gamma) \\\E c({X_P}) \le \Gamma  }  } V_P, \\
	V_{\min} &:= \min_{\substack{ P: I(P,W)=C_\varepsilon(\Gamma) \\ \E c({X_P}) \le \Gamma}} V_P
\end{align}
{by using $V_P$ instead of $V_{P, \theta_0}$.}
Formula \eqref{eq:second_order_formula} is due to Hayashi \cite{Hayashi2009} (with cost constraint), Polyanskiy, Poor, and Verd\'u \cite{PPV2010} (without cost constraints), and Strassen \cite{Strassen62}  (without cost constraints under the maximum error probability criterion). 
\QED
\end{e_rema}

\smallskip
Similarly to the first-order coding theorem, Theorem \ref{theo:general_mixed_DMC_achievable_2nd} reduces to the following corollary in the case where there are no cost constraints.
 
\begin{e_coro} \label{coro:general_mixed_DMC_achievable_2nd}
Let $\vect{W}$ be a {general} mixed memoryless channel with {measure} $w$.
For $\varepsilon \in [0,1)$ {and $R \ge0$}, {it holds that}
\begin{align}
{D_\varepsilon(R)}  \ge  \sup_{ P } \sup \left\{ S \, \Big| \, G_w(R, S|{P})  \le  \varepsilon  \right\}. \label{eq:2nd_order_general_mixture_no_cost_constraint}
\end{align}
\QED
\end{e_coro}


\section{Proofs of {Theorems \ref{theo:general_mixed_DMC_e-cap} and \ref{theo:general_mixed_DMC_achievable_2nd}}} \label{sect:proof}

\subsection{Lemmas} \label{sect:lemma}

We state several lemmas which are used to prove Theorems \ref{theo:general_mixed_DMC_e-cap} and \ref{theo:general_mixed_DMC_achievable_2nd}.
We first provide error bounds for {codes} of fixed length, which hold for any general {channel}.

\begin{e_lem}[{Feinstein's} Upper Bound \cite{Feinstein54}] \label{lem:Feinstein_bound}
For any input variable $X^n$  {with values in} $\mathcal{X}^n$, there exists an $(n, M_n, \varepsilon_n)$ code such that
\begin{align}
\varepsilon_n \le \Pr \left\{ \frac{1}{n} \log \frac{W^n(Y^n|X^n)}{P_{Y^n}(Y^n)} \le \frac{1}{n} \log M_n + \eta \right\} + e^{-n\eta}, \label{eq:Feinstein_bound}
\end{align}
where\footnote{For random variables $U$ and $V$, we let $P_U$ denote the probability distribution of $U$ and $P_{U|V}$ denote the conditional probability distribution of $U$ given $V$.} {$Y^n$ is the output variable due to $X^n$ via channel $W^n$ and}  $\eta > 0$ is an arbitrary positive number.
\QED
\end{e_lem}

{The following lemma was first established in \cite[Lemma 4]{Hayashi-Nagaoka2003} in the context of quantum channel coding. The proof for the classical version is stated in \cite[Sect.\ IX-B]{Hayashi2009}\footnote{{Later, we shall generalize this lemma to the mixed channel consisting of general component channels in Lemma \ref{lem:mixed_lower_bound2}, whose poof is given in Appendix \ref{append:proof_of_new_lower_bound}}.}.}
\begin{e_lem}[{Hayashi-Nagaoka's} Lower Bound \cite{Hayashi-Nagaoka2003}] \label{lem:HN_bound}
Let ${Q^n}$ be an arbitrary probability {distribution} on $\mathcal{Y}^n$.
 Every $(n, M_n, \varepsilon_n)$ code $\Code$ satisfies 
\begin{align}
\varepsilon_n \ge \Pr \left\{ \frac{1}{n} \log \frac{W^n(Y^n|X^n)}{{Q^n}(Y^n)} \le \frac{1}{n} \log M_n - \eta \right\} - e^{-n\eta}, \label{eq:HN_bound}
\end{align}
where {$X^n$ denotes the random variable subject to the uniform distribution on $\Code$, $Y^n$ denotes the output variable due to $X^n$ via channel $W^n$}, and $\eta > 0$ is an arbitrary {positive number}.
\QED
\end{e_lem}

\medskip
We next state lemmas for mixed channels. {We first arrange a so-called \emph{expurgated parameter space} which possesses a {useful} property and is still  asymptotically dominant {over} the whole parameter space.} 
Given a set of arbitrary i.i.d.\ product probability distributions ${Q_\theta^n = Q_\theta^{\times n}} $ on $\mathcal{Y}^n$, let ${Q^n}$ be
given as  
\begin{align}
{Q^n}(\vect{y}) := \int_{{\Theta}} {Q_\theta^n}(\vect{y}) dw(\theta)~~(\forall \vect{y} \in \mathcal{Y}^n), \label{eq:output_dist_set}
\end{align}
and define 
\begin{align}
{\Theta}(\vect{y}) := \left\{ \theta \in {\Theta} \,|\, {Q_\theta^n}(\vect{y}) \le e^{\sqrt[4]{n}} {Q^n}(\vect{y}) \right\}~~(\forall \vect{y} \in \mathcal{Y}^n)
\end{align}
and 
\begin{align}
\tilde{{\Theta}}(\vect{x},\vect{y}) := \left\{ \theta \in {\Theta} \,|\, W^n_\theta(\vect{y}|\vect{x}) \le  e^{\sqrt[4]{n}}  W^n(\vect{y}| \vect{x}) \right\}~~(\forall (\vect{x}, \vect{y}) \in \mathcal{X}^n \times \mathcal{Y}^n).
\end{align}
Let $S_k, k=1,2,\cdots, N_n,$ denote a type (empirical distribution) on $\mathcal{Y}^n$, where $N_n$ is the number of {all} distinct types.
Let $\tilde{S}_k, k=1,2,\cdots, \tilde{N}_n,$ denote a joint type on $\mathcal{X}^{{n}} \times \mathcal{Y}^n$, where $\tilde{N}_n$ is the number of {all} distinct joint types.
{Since $Q_\theta^n$ is an i.i.d. {product} probability distribution, the} subset ${\Theta}(\vect{y}) $ depends only on the type $S_k$ of $\vect{y}$, and therefore it {can} be denoted {as} ${\Theta}(S_k)$ {instead of ${\Theta}(\vect{y})$}. 
Likewise, {since $W_\theta^n (\vect{y}|\vect{x})$ is stationary and memoryless,}  the subset $\tilde{{\Theta}}(\vect{x}, \vect{y}) $ depends only on the joint type $\tilde{S}_k$ of $(\vect{x}, \vect{y})$, and therefore it {can} be denoted {as} $\tilde{{\Theta}}(\tilde{S}_k)$ {instead of $\tilde{{\Theta}}(\vect{x}, \vect{y})$}.
Using 
\begin{align}
{\Theta}_n := \bigcap_{k=1}^{N_n} {\Theta}(S_k)~~\mathrm{and}~~\tilde{{\Theta}}_n := \bigcap_{k=1}^{\tilde{N}_n} \tilde{{\Theta}}(\tilde{S}_k),
\end{align}
we define another set
\begin{align}
{\Theta}_n^* := {\Theta}_n  \cap \tilde{{\Theta}}_n.   \label{eq:Phi_n^*}
\end{align}

\begin{e_lem}\label{lem:dominant_component}
Let $\vect{W}$ be a {general} mixed memoryless channel with {measure} $w$.
{Given a set of arbitrary i.i.d.\ product probability distributions ${Q_\theta^n} $ on $\mathcal{Y}^n$, let ${Q^n}$ be 
defined by \eqref{eq:output_dist_set}.}
Then,  it holds that 
\begin{align}
\int_{{\Theta}^*_n} dw(\theta) \ge 1 - 2(n+1)^{|\mathcal{X}| \cdot |\mathcal{Y}|} e^{-\sqrt[4]{n}}.  \label{eq:dominant_subset}
\end{align}
\end{e_lem}
(Proof)~~See Appendix \ref{append:proof_lem_dominant_component}.
\QED

\medskip
The following lemmas play a key role in proving the coding theorems for mixed channels.
 \begin{e_lem}[{Upper Decomposition Lemma}] \label{lem:mixed_upper_decomposition}
 Let $\vect{W}$ be a {general} mixed memoryless channel with {measure} $w$.
 Then, it holds that 
\begin{align}
\Pr \left\{ {\frac{1}{n}} \log \frac{W^n(Y_\theta^n|X^n)}{P_{Y^n}(Y_\theta^n)} \le z_n \right\} \le \Pr \left\{ {\frac{1}{n}} \log \frac{W_\theta^n(Y_\theta^n|X^n)}{P_{Y_\theta^n}(Y_\theta^n)} \le z_n + {{\frac{\gamma}{\sqrt{n}}} + {\frac{1}{\sqrt[4]{n^3}}} } \right\} + e^{-\sqrt{n}\gamma} \nonumber \\
~~~(\forall \theta \in {\Theta}_n^*), \label{eq:mixed_upper_bound}
\end{align}
where $\gamma > 0$ and $z_n>0$ are arbitrary numbers, {and $Y_\theta^n$ indicates the output variable due to the input $X^n$ via channel $W_\theta^n$}.
\end{e_lem}
(Proof)~~See Appendix \ref{append:proof_U_decomposition_lemma}.
\QED

\medskip
 \begin{e_lem}[{Lower Decomposition Lemma}] \label{lem:mixed_lower_decomposition}
Let $\vect{W}$ be a {general} mixed memoryless channel with {measure} $w$.
Given a set of arbitrary i.i.d.\ product probability distributions $Q_\theta^n$ on $\mathcal{Y}^n$, let ${Q^n}$ be 
defined by \eqref{eq:output_dist_set}.
 Then, it holds that
\begin{align}
\Pr \left\{ {\frac{1}{n}} \log \frac{W^n(Y_\theta^n|X^n)}{{Q^n}(Y_\theta^n)} \le z_n \right\} \ge \Pr \left\{ {\frac{1}{n}} \log \frac{W_\theta^n(Y_\theta^n|X^n)}{{Q_\theta^n}(Y_\theta^n)} \le z_n - {\frac{\gamma}{\sqrt{n}}} - {\frac{1}{\sqrt[4]{n^3}}} \right\} - e^{-\sqrt{n}\gamma} \nonumber \\
~~~(\forall \theta \in {\Theta}_n^*), \label{eq:mixed_lower_bound}
\end{align}
where $\gamma > 0$ and $z_n>0$ are arbitrary numbers, {and $Y_\theta^n$ indicates the output variable due to the input $X^n$ via channel $W_\theta^n$}.
\end{e_lem}
(Proof)~~See Appendix \ref{append:proof_L_decomposition_lemma}.
\QED

{
\begin{e_rema} \label{rema:proof_duality}
As we shall show in the proof of Theorem \ref{theo:general_mixed_DMC_e-cap} in the next subsection,  there exists {an} interesting duality between the achievability proof and the converse proof based on Lemmas \ref{lem:mixed_upper_decomposition} and \ref{lem:mixed_lower_decomposition}. 
Using Upper/Lower Decomposition Lemma has been {the} standard technique in the analysis of the optimum coding rate in various problems in information theory such as source coding {\cite[Sect.\ 1.4]{Han2003}}, \cite{Nomura-Han2014}, random number generation \cite{Nomura-Han2013}, and hypothesis testing {\cite[Sect.\ 4.2]{Han2003}} for mixed sources.
The proof of Theorem \ref{theo:general_mixed_DMC_e-cap} demonstrates that we may also use this standard technique for mixed memoryless channels.
Later, we shall also demonstrate in Sect.\ \ref{sect:proof_WO_mixed_DMC_2nd} that Lemma \ref{lem:mixed_lower_bound2} can be used as a powerful alternative to Lemmas \ref{lem:HN_bound} and \ref{lem:mixed_lower_decomposition}, and it saves several steps of the converse proof.
\QED
\end{e_rema}}

\subsection{Proof of Theorem \ref{theo:general_mixed_DMC_e-cap}} \label{sect:proof_general_mixed_DMC_e-cap}

\noindent
(Proof of Direct Part)

{Define}
\begin{align}
{\overline{C}_\varepsilon(\Gamma)} := \sup_{{P}: {\E c({X_P}) \le \Gamma}} \sup \left\{ R \, \Big| \, \int_{\{\theta | I {(}{P},W_\theta) < R\}} dw(\theta) \le \varepsilon \right\}, \label{eq:C_epsilon}
\end{align}
{and then} for any {small} $\delta >0$ there exists an input distribution ${P_0} \in \PX$ {such that $\E c({X_{P_0}}) \le \Gamma$ and}
\begin{align}
\sup \left\{ R \, \Big| \, \int_{\{\theta | I({P_0},W_\theta) < R\}} dw(\theta) \le \varepsilon \right\} \ge {\overline{C}_\varepsilon(\Gamma)} - \delta. \label{eq:ineq1}
\end{align}
We fix such a ${P_0}$ {and show that
\begin{align}
R = {\overline{C}_\varepsilon(\Gamma)} - 4 \delta. \label{eq:rate_set2}
\end{align}
is {$(\varepsilon|\Gamma)$}-achievable.}

{
Without loss of generality, we assume that {the} elements in {$\mathcal{X}=\{1, 2, \ldots, |\mathcal{X}| \}$} are indexed {so} that 
$c(1) \ge c(2) \ge \cdots \ge c(|\mathcal{X}|)$.
We define the type $P_n$ {on $\mathcal{X}^n$}  {so} that
\begin{align}
P_n(x) &= \frac{\lfloor n P_0(x) \rfloor}{n}~~(x=1,2,\ldots, |\mathcal{X}|-1), \\
P_n(|\mathcal{X}|) &= 1 - \sum_{x=1}^{|\mathcal{X}|-1} P_n(x).
\end{align}
Then, it is readily shown that
\begin{align}
\sum_{x \in \mathcal{X}} P_n(x) c(x) &\le \Gamma,  \label{eq:cost_satisfy}\\
|P_n(x)-P_0(x)| &\le \frac{|\mathcal{X}|}{n}~~(\forall x \in \mathcal{X}), \label{eq:type_diff}
\end{align}
and 
\begin{align}
\lim_{n \rightarrow \infty} P_n(x) = P_0(x)~~(\forall x \in \mathcal{X}), \label{eq:type_convergence}
\end{align}
{where \eqref{eq:cost_satisfy} follows because $P_0$ satisfies $\sum_{x \in \mathcal{X}} P_0(x) c(x) \le \Gamma$.}
}

{Let $T_n$ be the set of all sequences $\vect{x} \in \mathcal{X}^n$ {of} type $P_n$, and consider the input random variable $X^n$ uniformly distributed on $T_n$. }
Using Lemma \ref{lem:Feinstein_bound} with $\frac{1}{n} \log M_n = R$ and $\eta = \frac{\gamma}{\sqrt{n}}$, {where $\gamma > 0$ is an arbitrary positive number}, we obtain the following chain of expansions
\begin{align}
\limsup_{n \rightarrow \infty} \varepsilon_n 
&\le \limsup_{n \rightarrow \infty} \Pr \left\{ \frac{1}{n} \log \frac{W^n(Y^n|X^n)}{P_{Y^n}(Y^n)} \le R + \frac{\gamma}{\sqrt{n}} \right\} \nonumber \\
&= \limsup_{n \rightarrow \infty} \int_{{\Theta}} dw(\theta) \Pr \left\{ \frac{1}{n} \log \frac{W^n(Y_\theta^n|X^n)}{P_{Y^n}(Y_\theta^n)} \le R + \frac{\gamma}{\sqrt{n}} \right\} \nonumber \\
&= \limsup_{n \rightarrow \infty} \left[ \int_{{\Theta}_n^*} dw(\theta) \Pr \left\{ \frac{1}{n} \log \frac{W^n(Y_\theta^n|X^n)}{P_{Y^n}(Y_\theta^n)} \le R + \frac{\gamma}{\sqrt{n}} \right\} \right. \nonumber \\
&~~~~~~~~~~~\left. + \int_{{\Theta} - {\Theta}_n^*} dw(\theta) \Pr \left\{ \frac{1}{n} \log \frac{W^n(Y_\theta^n|X^n)}{P_{Y^n}(Y_\theta^n)} \le R + \frac{\gamma}{\sqrt{n}} \right\} \right] \nonumber \\
&\le \limsup_{n \rightarrow \infty}\int_{{\Theta}_n^*} dw(\theta) \Pr \left\{ \frac{1}{n} \log \frac{W^n(Y_\theta^n|X^n)}{P_{Y^n}(Y_\theta^n)} \le R + \frac{\gamma}{\sqrt{n}} \right\}  \nonumber \\
&~~~~+ \limsup_{n \rightarrow \infty} \int_{{\Theta} - {\Theta}_n^*} dw(\theta) \Pr \left\{ \frac{1}{n} \log \frac{W^n(Y_\theta^n|X^n)}{P_{Y^n}(Y_\theta^n)} \le R + \frac{\gamma}{\sqrt{n}} \right\}  \nonumber \\
&\le \limsup_{n \rightarrow \infty}\int_{{\Theta}_n^*} dw(\theta) \Pr \left\{ \frac{1}{n} \log \frac{W^n(Y_\theta^n|X^n)}{P_{Y^n}(Y_\theta^n)} \le R + \frac{\gamma}{\sqrt{n}} \right\} + \limsup_{n \rightarrow \infty} \int_{{\Theta} - {\Theta}_n^*} dw(\theta)  \nonumber \\
&= \limsup_{n \rightarrow \infty}\int_{{\Theta}_n^*} dw(\theta) \Pr \left\{ \frac{1}{n} \log \frac{W^n(Y_\theta^n|X^n)}{P_{Y^n}(Y_\theta^n)} \le R + \frac{\gamma}{\sqrt{n}} \right\}.  \label{ineq1}
\end{align}
Here, we have used
\begin{align}
\int_{{\Theta} - {\Theta}_n^*} dw(\theta) \le 2(n+1)^{|\mathcal{X}| \cdot |\mathcal{Y}|} e^{-\sqrt[4]{n}} \label{eq:non_dominant_component}
\end{align}
 (cf.\ Lemma \ref{lem:dominant_component}) to obtain \eqref{ineq1}.
We apply Lemma \ref{lem:mixed_upper_decomposition} with {$z_n = R + \frac{\gamma}{\sqrt{n}}$} to \eqref{ineq1} to obtain
\begin{align}
\limsup_{n \rightarrow \infty} \varepsilon_n 
&\le \limsup_{n \rightarrow \infty}\int_{{\Theta}_n^*} dw(\theta) \Pr \left\{ \frac{1}{n} \log \frac{W_\theta^n(Y_\theta^n|X^n)}{P_{Y_\theta^n}(Y_\theta^n)} \le R + {\frac{{2 \gamma}}{\sqrt{n}} + \frac{1}{\sqrt[4]{n^3}} } \right\} \nonumber \\
&\le {\limsup_{n \rightarrow \infty}}  \int_{{\Theta}} dw(\theta) \Pr \left\{ \frac{1}{n} \log \frac{W_\theta^n(Y_\theta^n|X^n)}{P_{Y_\theta^n}(Y_\theta^n)} \le R + {\frac{{2 \gamma}}{\sqrt{n}} + \frac{1}{\sqrt[4]{n^3}} } \right\}  \nonumber \\
&\le \int_{{\Theta}} dw(\theta) \limsup_{n \rightarrow \infty} \Pr \left\{ \frac{1}{n} \log \frac{W_\theta^n(Y_\theta^n|X^n)}{P_{Y_\theta^n}(Y_\theta^n)} \le R + {\frac{{2 \gamma}}{\sqrt{n}} + \frac{1}{\sqrt[4]{n^3}} } \right\},  \label{ineq2} 
\end{align}
where the inequality {in} \eqref{ineq2} is due to Fatou's lemma.
{
Now notice that
\begin{align}
P_{Y_\theta^n}(\vect{y}) &=  \frac{1}{|T_n|} \sum_{\vect{x} \in T_n} W_\theta^n(\vect{y}|\vect{x}) \nonumber \\
&\le (n+1)^{|\mathcal{X}|} \sum_{\vect{x} \in T_n}  e^{-nH(P_n)} W_\theta^n(\vect{y}|\vect{x}) \nonumber \\
&= (n+1)^{|\mathcal{X}|} \sum_{\vect{x} \in T_n}  \prod_{i=1}^n P_n(x_i) W_\theta(y_i|x_i) \nonumber \\
&=  (n+1)^{|\mathcal{X}|}  (P_nW_\theta)^{\times n}(\vect{y})~~(\forall \vect{y} \in \mathcal{Y}^n), \label{eq:ineq2c}
\end{align}
where $ (P_nW_\theta)^{\times n}$ denotes the {$n$} product distribution of 
\begin{align}
P_n W_\theta (y) := \sum_{x \in \mathcal{X}} P_n(x) W_\theta(y|x)~~(\forall y \in \mathcal{Y}).
\end{align}
Plugging inequality \eqref{eq:ineq2c} into  \eqref{ineq2}, we obtain
\begin{align}
\limsup_{n \rightarrow \infty} \varepsilon_n 
&\le \int_{{\Theta}} dw(\theta) \limsup_{n \rightarrow \infty} \Pr \left\{ \frac{1}{n} \log \frac{W_\theta^n(Y_\theta^n|X^n)}{ (P_nW_\theta)^{\times n} (Y_\theta^n)} \le R + {\frac{{2 \gamma}}{\sqrt{n}} + \frac{1}{\sqrt[4]{n^3} }} + \frac{|\mathcal{X}|}{n} \log(n+1) \right\}\nonumber \\
&\le \int_{{\Theta}} dw(\theta) \limsup_{n \rightarrow \infty} \Pr \left\{ \frac{1}{n} \log \frac{W_\theta^n(Y_\theta^n|X^n)}{ (P_nW_\theta)^{\times n} (Y_\theta^n)} \le R + \delta \right\}. \label{ineq2d} 
\end{align}
Inequality \eqref{ineq2d} implies that there exists $\vect{x}_n \in \mathcal{X}^n$ {of} type $P_n$ such that
\begin{align}
\limsup_{n \rightarrow \infty} \varepsilon_n 
&\le \int_{{\Theta}} dw(\theta) \limsup_{n \rightarrow \infty} \Pr \left\{ \frac{1}{n} \log \frac{W_\theta^n(Y_\theta^n|\vect{x}_n)}{ (P_nW_\theta)^{\times n} (Y_\theta^n)} \le R + \delta \Big| X^n = \vect{x}_n \right\} \label{ineq2e} 
\end{align}
}
{Now, we can write as
\begin{align}
 \frac{1}{n} \log \frac{W_\theta^n(Y_\theta^n|\vect{x}_n)}{ (P_nW_\theta)^{\times n} (Y_\theta^n)}  =  \frac{1}{n} \sum_{i=1}^n \log \frac{W_\theta(Y_{\theta, i}|x_i)}{P_nW_\theta(Y_{\theta,i})}, \label{eq:independent_sum}
\end{align}
where 
\begin{align}
\vect{x}_n &= (x_1, x_2, \cdots, x_n), \nonumber \\
Y_\theta^n &=(Y_{\theta, 1}, Y_{\theta, 2}, \cdots, Y_{\theta, n}). \nonumber 
\end{align}
Notice here that $Y_{\theta, 1}, Y_{\theta, 2}, \ldots, Y_{\theta, n}$ are conditionally independent random variables given $X^n = \vect{x}_n$ (under the conditional distribution $W^n_\theta(\cdot | \vect{x}_n)$), and therefore the right-hand side of \eqref{eq:independent_sum}  is a sum of conditionally independent random variables given $X^n = \vect{x}_n$ with conditional mean
\begin{align}
 \E \left\{\frac{1}{n} \sum_{i=1}^n \log \frac{W_\theta(Y_{\theta, i}|x_i)}{P_nW_\theta(Y_{\theta,i})} \Big| \, X^n = \vect{x}_n \right\} & =I(P_n, W_\theta) \label{eq:independent_sum_mean}
\end{align}
and conditional variance
\begin{align}
&\V \left\{ \frac{1}{n} \sum_{i=1}^n \log \frac{W_\theta(Y_{\theta, i}|x_i)}{P_nW_\theta(Y_{\theta,i})} \Big| \, X^n = \vect{x}_n  \right\} \nonumber \\
 &~~~~~~~~~~~~ = \sum_{x \in \mathcal{X}} \sum_{y \in \mathcal{Y}} P_n(x) W_\theta (y|x) \left( \log \frac{W_\theta(y|x)}{P_nW_\theta(y)} - D(W_\theta (\cdot | x) || P_n W_\theta) \right)^2 \nonumber \\
 &~~~~~~~~~~~~ = {V_{\theta, P_n}}. \label{eq:independent_sum_variance}
\end{align}}

Then, we can invoke the weak law of large numbers to the probability term $\Pr\{ \cdot \}$ in {\eqref{ineq2e}}.
{To do so,} we split {the parameter space} ${\Theta}$ as follows:
\begin{align}
{\Theta}_1 &:= \{ \theta \in {\Theta} | I({P_0}, W_\theta) < R + {\delta} \}, \label{eq:Phi_1} \\
{\Theta}_2 &:= \{ \theta \in {\Theta} | I({P_0}, W_\theta) = R + {\delta} \}, \label{eq:Phi_2} \\
{\Theta}_3 &:= \{ \theta \in {\Theta} | I({P_0}, W_\theta) > R + {\delta} \}. \label{eq:Phi_3}
\end{align}
It is easily verified that
\begin{align}
\limsup_{n \rightarrow \infty} \Pr \left\{ \frac{1}{n} \log \frac{W_\theta^n(Y_\theta^n|\vect{x}_n)}{(P_n W_\theta)^{\times n}(Y_\theta^n)} \le R + {\delta} \Big| X^n = \vect{x}_n \right\} = \left\{
\begin{array}{cc}
1, & \mathrm{if}~\theta \in {\Theta}_1 \\ 
0, & \mathrm{if}~\theta \in {\Theta}_3
\end{array}
\right.   \label{ineq3}
\end{align}
by virtue of the weak law of large numbers and \eqref{eq:type_convergence}, {where we should notice that the inequality
\begin{align}
\max_{P} V_{\theta, P} < + \infty ~~(\forall \theta \in \Theta) \label{eq:bounded_variance}
\end{align}
holds due to Han \cite[Remark 3.1.1]{Han2003} and Polyanskiy et al. \cite[Lemma 62]{PPV2010}.}
Then, {\eqref{ineq2e}} is {rewritten} as
\begin{align}
\limsup_{n \rightarrow \infty} \varepsilon_n 
&\le \int_{{\Theta}_1 \cup {\Theta}_2} dw(\theta) = \int_{\{ \theta | I({P_0},W_\theta) \le {{\overline{C}_\varepsilon(\Gamma)} - 3 \delta} \}} dw(\theta) \nonumber \\
 &\le  \int_{\{ \theta | I({P_0},W_\theta)  {\, < {\overline{C}_\varepsilon(\Gamma)} - 2 \delta} \}} dw(\theta) \le \varepsilon, \label{ineq4}
\end{align}
where the last inequality follows from {\eqref{eq:ineq1}}.
Hence, $R = {\overline{C}_\varepsilon(\Gamma)} - {4 \delta}$ is {$(\varepsilon|\Gamma)$-achievable}.
\QED

\medskip
\noindent
(Proof of Converse Part)

Assume that {$R$ is $(\varepsilon|\Gamma)$-achievable}. 
{By the definition of {$(\varepsilon|\Gamma)$-achievable} rates, there exists an $(n, M_n, \varepsilon_n)$ code $\Code$ {with cost constraint $\Gamma$} such that, {for an arbitrary $\delta>0$},
\begin{align}
\frac{1}{n} \log M_n \ge R - \delta ~~(\forall n > n_0). \label{eq:rate_cond_conv}
\end{align}
} 
By Lemma \ref{lem:HN_bound} {with $\eta = \frac{\gamma}{\sqrt{n}}$}, {any} $(n, M_n, \varepsilon_n)$ code {$\Code$} satisfies
\begin{align}
\varepsilon_n \ge \Pr \left\{ \frac{1}{n} \log \frac{W^n(Y^n|X^n)}{{Q^n}(Y^n)} \le \frac{1}{n} \log M_n - \frac{\gamma}{\sqrt{n}} \right\} - e^{-\sqrt{n}\gamma}, \label{eq:HN_bound2}
\end{align}
where {$X^n$ denotes the random variable subject to} the uniform distribution on the code $\Code$ and $\gamma > 0$ is an arbitrary number.
The output distribution ${Q^n}$ in \eqref{eq:HN_bound2} is set as follows:
Letting ${Q_\theta^n}$ be an output distribution on $\mathcal{Y}^n$ indexed by $\theta \in {\Theta}$ {such as}
\begin{align}
{Q_\theta^n} (\vect{y}) := \frac{1}{{N_n}} \sum_{P_n \in \mathcal{T}_n} (P_nW_\theta)^{\times n}(\vect{y})~~(\forall \theta \in {\Theta}, \forall \vect{y} \in \mathcal{Y}^n) \label{eq:output_dist}
\end{align}
where {$\mathcal{T}_n$ denotes the set of all types on $\mathcal{X}^n$ of size  ${N_n} := |\mathcal{T}_n|$}, and $(P_nW_\theta)^{\times n}$ denotes the {$n$} product distribution of {$P_nW_\theta$}.
Using {this} $\{{Q_\theta^n}\}_{\theta \in {\Theta}}$, {we define ${Q^n}$ as}
\begin{align}
{Q^n}(\vect{y}) := \int_{{\Theta}} {Q_\theta^n} (\vect{y}) dw(\theta)~~(\forall \vect{y} \in \mathcal{Y}^n), \label{eq:output_dist2}
\end{align}
{where we notice that} $Q_\theta^n(\vect{y})$ depends only on the type of $\vect{y}$, {and} so does ${Q^n (\vect{y})}$.  

Since $R$ is {$(\varepsilon|\Gamma)$-achievable}, the following expansion follows from \eqref{eq:rate_cond_conv} and \eqref{eq:HN_bound2}:
\begin{align}
\varepsilon 
& \ge \limsup_{n \rightarrow \infty} \Pr \left\{ \frac{1}{n} \log \frac{W^n(Y^n|X^n)}{{Q^n}(Y^n)} \le {R - \delta} - \frac{\gamma}{\sqrt{n}} \right\} \nonumber \\
& \ge \limsup_{n \rightarrow \infty} \int_{{\Theta}} dw(\theta) \Pr \left\{ \frac{1}{n} \log \frac{W^n(Y_\theta^n|X^n)}{{Q^n}(Y_\theta^n)} \le {R - 2 \delta}   \right\} \nonumber \\
& \ge \limsup_{n \rightarrow \infty} \int_{{\Theta}_n^*} dw(\theta) \Pr \left\{ \frac{1}{n} \log \frac{W^n(Y_\theta^n|X^n)}{{Q^n}(Y_\theta^n)} \le {R - 2 \delta}    \right\}. \label{eq:HN_bound3}
\end{align}
Applying Lemma \ref{lem:mixed_lower_decomposition} {with $z_n = {R - 2\delta}$} to \eqref{eq:HN_bound3} yields
\begin{align}
\varepsilon 
& \ge \limsup_{n \rightarrow \infty} \int_{{\Theta}_n^*} dw(\theta) \Pr \left\{ \frac{1}{n} \log \frac{W_\theta^n(Y_\theta^n|X^n)}{{Q_\theta^n}(Y_\theta^n)} \le {R - 2 \delta}  - {\frac{\gamma}{\sqrt{n}} - { \frac{1}{\sqrt[4]{n^3}}} }  \right\} \nonumber \\
& \ge \limsup_{n \rightarrow \infty} \int_{{\Theta}_n^*} dw(\theta) \Pr \left\{ \frac{1}{n} \log \frac{W_\theta^n(Y_\theta^n|X^n)}{{Q_\theta^n}(Y_\theta^n)} \le {R - 3 \delta}  \right\} \nonumber \\
& {=} \limsup_{n \rightarrow \infty} \left[ \int_{{\Theta}} dw(\theta) \Pr \left\{ \frac{1}{n} \log \frac{W_\theta^n(Y_\theta^n|X^n)}{{Q_\theta^n}(Y_\theta^n)} \le {R - 3 \delta} \right\} \right.  \nonumber \\
& ~~~~~~~~~~~~\left. - \int_{{\Theta} - {\Theta}_n^*} dw(\theta) \Pr \left\{ \frac{1}{n} \log \frac{W_\theta^n(Y_\theta^n|X^n)}{{Q_\theta^n}(Y_\theta^n)} \le {R - 3 \delta} \right\}\right] \nonumber \\
& \ge \limsup_{n \rightarrow \infty}   \int_{{\Theta}} dw(\theta) \Pr \left\{ \frac{1}{n} \log \frac{W_\theta^n(Y_\theta^n|X^n)}{{Q_\theta^n}(Y_\theta^n)} \le {R - 3 \delta} \right\}    \nonumber \\
& ~~ - \limsup_{n \rightarrow \infty} \int_{{\Theta} - {\Theta}_n^*} dw(\theta) \Pr \left\{ \frac{1}{n} \log \frac{W_\theta^n(Y_\theta^n|X^n)}{{Q_\theta^n}(Y_\theta^n)} \le {R - 3 \delta} \right\}  \nonumber \\
& \ge \limsup_{n \rightarrow \infty}   \int_{{\Theta}} dw(\theta) \Pr \left\{ \frac{1}{n} \log \frac{W_\theta^n(Y_\theta^n|X^n)}{{Q_\theta^n}(Y_\theta^n)} \le{R - 3 \delta} \right\}  - \limsup_{n \rightarrow \infty} \int_{{\Theta} - {\Theta}_n^*} dw(\theta)  \nonumber \\
& = \limsup_{n \rightarrow \infty}   \int_{{\Theta}} dw(\theta) \Pr \left\{ \frac{1}{n} \log \frac{W_\theta^n(Y_\theta^n|X^n)}{{Q_\theta^n}(Y_\theta^n)} \le {R - 3 \delta} \right\}.   \label{eq:HN_bound4} 
\end{align}
Here, we {have used} \eqref{eq:non_dominant_component} to obtain \eqref{eq:HN_bound4}.
Notice {that}
\begin{align}
& \int_{{\Theta}} dw(\theta) \Pr \left\{ \frac{1}{n} \log \frac{W_\theta^n(Y_\theta^n|X^n)}{{Q_\theta^n}(Y_\theta^n)} \le {R - 3 \delta} \right\} \nonumber \\
&~~~~~~~ = \sum_{\vect{x} \in \mathcal{C}_n}  \frac{1}{M_n} \int_{{\Theta}} dw(\theta) \Pr \left\{ \frac{1}{n} \log \frac{W_\theta^n(Y_\theta^n|\vect{x})}{{Q_\theta^n}(Y_\theta^n)} \le {R - 3 \delta} \Big| X^n = \vect{x} \right\},  \label{eq:HN_bound5} 
\end{align}
 and therefore there {exists} a codeword $\vect{x}_n \in \mathcal{C}_n$ such that
\begin{align}
&\int_{{\Theta}} dw(\theta) \Pr \left\{ \frac{1}{n} \log \frac{W_\theta^n(Y_\theta^n|X^n)}{{Q_\theta^n}(Y_\theta^n)} \le {R - 3 \delta} \right\} \nonumber \\
&~~~~\ge  \int_{{\Theta}} dw(\theta) \Pr \left\{ \frac{1}{n} \log \frac{W_\theta^n(Y_\theta^n|\vect{x}_n)}{{Q_\theta^n}(Y_\theta^n)} \le {R - 3 \delta} \Big| X^n = \vect{x}_n \right\} ~~~(\forall n > n_0).  \label{eq:HN_bound6} 
\end{align}
{Let $P_n$ denote the type of such $\vect{x}_n$}. 
{By} \eqref{eq:output_dist}, the right-hand side of \eqref{eq:HN_bound6} can be lower bounded as
\begin{align}
&\int_{{\Theta}} dw(\theta) \Pr \left\{ \frac{1}{n} \log \frac{W_\theta^n(Y_\theta^n|\vect{x}_n)}{{Q_\theta^n}(Y_\theta^n)} \le {R - 3 \delta} \Big| X^n = \vect{x}_n \right\} \nonumber \\
&~~~~~~~~~~~\ge \int_{{\Theta}} dw(\theta) \Pr \left\{ \frac{1}{n} \log \frac{W_\theta^n(Y_\theta^n|\vect{x}_n)}{(P_nW_\theta)^{\times n}(Y_\theta^n)} \le {R - 3 \delta} - \frac{\log N_n}{n}  \Big| X^n = \vect{x}_n \right\} \nonumber \\
&~~~~~~~~~~~\ge \int_{{\Theta}} dw(\theta) \Pr \left\{ \frac{1}{n} \log \frac{W_\theta^n(Y_\theta^n|\vect{x}_n)}{(P_nW_\theta)^{\times n}(Y_\theta^n)} \le {R - 4 \delta} \Big| X^n = \vect{x}_n \right\}~~~(\forall n \ge \tilde{n}_0),  \label{eq:HN_bound7} 
\end{align}
where we {have used} the inequality $N_n \le (n+1)^{|\mathcal{X}|}$ to obtain \eqref{eq:HN_bound7}.
Combining \eqref{eq:HN_bound4}, {\eqref{eq:HN_bound6}}, and \eqref{eq:HN_bound7} yields
\begin{align}
\varepsilon 
& \ge \limsup_{n \rightarrow \infty}  \int_{{\Theta}} dw(\theta) \Pr \left\{ \frac{1}{n} \log \frac{W_\theta^n(Y_\theta^n|\vect{x}_n)}{(P_nW_\theta)^{\times n}(Y_\theta^n)} \le {R - 4 \delta} \Big| X^n = \vect{x}_n \right\}. \label{eq:HN_bound8}
\end{align}

Since $\{P_n\}_{n > {\tilde{n}_0}}$ is a sequence in $\PX$ (compact set), it always contains a converging {subsequence} $\{P_{n_1}, P_{n_2}, \cdots \}$, where $n_1 < n_2< \cdots < \rightarrow \infty$. We denote the {convergent} point by $P_0$;
\begin{align}
\lim_{i \rightarrow \infty} P_{n_i} = P_0,
\end{align}
{where it should be noticed that $P_0$ satisfies cost constraint: $\E c(X_{P_0}) \le \Gamma$ because $P_n$ satisfies the same cost constraint $\Gamma$.}
For {notational} simplicity, we relabel $n_k$ as {$m = n_1, n_2,\cdots$}.
Then, {in view of 
\begin{align}
&\limsup_{n \rightarrow \infty}  \int_{{\Theta}} dw(\theta) \Pr \left\{ \frac{1}{n} \log \frac{W_\theta^n(Y_\theta^n|\vect{x}_n)}{(P_nW_\theta)^{\times n}(Y_\theta^n)} \le {R - 4 \delta} \Big| X^n = \vect{x}_n \right\} \nonumber \\
&~~~\ge \limsup_{m \rightarrow \infty}  \int_{{\Theta}} dw(\theta) \Pr \left\{ \frac{1}{m} \log \frac{W_\theta^m(Y_\theta^m|\vect{x}_m)}{(P_mW_\theta)^{\times m}(Y_\theta^m)} \le {R - 4 \delta} \Big| X^m = \vect{x}_m \right\},
\end{align}
}%
 \eqref{eq:HN_bound8} becomes
\begin{align}
\varepsilon 
& \ge \limsup_{{{m}} \rightarrow \infty}  \int_{{\Theta}} dw(\theta) \Pr \left\{ \frac{1}{{{m}}} \log \frac{W_\theta^{{m}}(Y_\theta^{{m}}|\vect{x}_m)}{(P_{{m}}W_\theta)^{\times {{m}}}(Y_\theta^{{m}})} \le {R - 4 \delta} \Big| X^{{m}} = \vect{x}_{{m}} \right\} \nonumber \\
& \ge \int_{{\Theta}} dw(\theta)  \liminf_{{{m}} \rightarrow \infty}   \Pr \left\{ \frac{1}{{{m}}} \log \frac{W_\theta^{{m}}(Y_\theta^{{m}}|\vect{x}_{{m}})}{(P_{{m}}W_\theta)^{\times {{m}}}(Y_\theta^{{m}})} \le {R - 4 \delta} \Big| X^{{m}} = \vect{x}_{{m}} \right\} \label{eq:HN_bound8b} \\
& \ge \int_{{\Theta}_1} dw(\theta)  \liminf_{{{m}} \rightarrow \infty}   \Pr \left\{ \frac{1}{{{m}}} \log \frac{W_\theta^{{m}}(Y_\theta^{{m}}|\vect{x}_{{m}})}{(P_{{m}}W_\theta)^{\times {{m}}}(Y_\theta^{{m}})} \le {R - 4 \delta} \Big| X^{{m}} = \vect{x}_{{m}} \right\}, \label{eq:HN_bound8c}
\end{align}
where the inequality in \eqref{eq:HN_bound8b} is due to Fatou's lemma, and ${\Theta}_1$ is defined as
\begin{align}
{\Theta}_1 &:= \{ \theta \in {\Theta} | I(P_0, W_\theta) < {R - 4 \delta} \}.
\end{align}
{Set $\vect{x}_{{m}} = (x_1,x_2,\cdots,x_m)$, and then}
\begin{align}
\frac{1}{{m}} \log \frac{W_\theta^{{m}}(Y_\theta^{{m}}|\vect{x}_{{m}})}{(P_{{m}}W_\theta)^{\times {{m}}}(Y_\theta^{{m}})} 
= \frac{1}{{m}}  \sum_{i=1}^{{m}} \log \frac{W_\theta(Y_{\theta, i}|{x_i})}{(P_{{m}}W_\theta)(Y_{\theta, i})} 
\end{align}
is a sum of {conditionally} independent random variables given $X^m = \vect{x}_m$, and its expectation and variance under $W_\theta^{{m}}(\cdot|\vect{x}_{{m}})$ are given by
\begin{align}
 \E \left\{ \frac{1}{{m}} \log \frac{W_\theta^{{m}}(Y_\theta^{{m}}|\vect{x}_{{m}})}{(P_{{m}}W_\theta)^{\times {{m}}}(Y_\theta^{{m}})}   {\Big| \, X^m = \vect{x}_m } \right\}
= I(P_{{m}}, W_\theta) \label{eq:expectation}
\end{align}
and 
\begin{align}
&\V \left\{ \frac{1}{{m}} \log \frac{W_\theta^{{m}}(Y_\theta^{{m}}|\vect{x}_{{m}})}{(P_{{m}}W_\theta)^{\times {{m}}}(Y_\theta^{{m}})}  {\Big| \, X^m = \vect{x}_m } \right\} \nonumber \\
&~~~ = \sum_{x \in \mathcal{X}} \sum_{y\in \mathcal{Y}} P_m(x) W_\theta(y|x) \left( \log \frac{W_\theta(y|x)}{(P_m W_\theta)(y)} - D\big(W_\theta(\cdot |x) || P_m W_\theta \big)\right)^2 \nonumber \\
&~~~= V_{\theta, P_m}, \label{eq:cond_variance}
\end{align}
{respectively}.
Hence, the weak law of large numbers guarantees
\begin{align}
 \liminf_{{{m}} \rightarrow \infty}   \Pr \left\{ \frac{1}{{{m}}} \log \frac{W_\theta^{{m}}(Y_\theta^{{m}}|\vect{x}_{{m}})}{(P_{{m}}W_\theta)^{\times {{m}}}(Y_\theta^{{m}})} \le {R - 4 \delta} \Big| X^{{m}} = \vect{x}_{{m}} \right\} = 1~~~(\forall \theta \in {\Theta}_1) .\label{ineq5}
\end{align}
{Thus,} \eqref{eq:HN_bound8c} is {rewritten} as
\begin{align}
 \varepsilon
&\ge \int_{{\Theta}_1} dw(\theta) = \int_{\{ \theta | I(P_0,W_\theta) {<} {R - 4 \delta} \}} dw(\theta). \label{ineq6}
\end{align}
{Therefore}, from the definition of ${\overline{C}_\varepsilon(\Gamma)}$ (cf.\ \eqref{eq:C_epsilon}), {we have}
\begin{align}
{R - 4 \delta \le \overline{C}_\varepsilon(\Gamma)}.
\end{align}
{On the other hand, since $\delta > 0 $ is arbitrary, we conclude that $R \le \overline{C}_\varepsilon(\Gamma)$.}
\QED

\subsection{Proof of Theorem \ref{theo:general_mixed_DMC_achievable_2nd}} \label{sect:proof_general_mixed_DMC_achievable_2nd}

We first define 
\begin{align}
{\overline{D}_\varepsilon(R|\Gamma)} := \sup_{P: {\E c({X_P}) \le \Gamma}} \sup \left\{S \,|\, G_w(R,S|P) \le \varepsilon  \right\}, \label{eq:overline_D_def}
\end{align}
where see \eqref{eq:func_Gw} as for the definition of $G_w(R,S|P)$.
Then, for any $\delta > 0$ there exists an input distribution $P_0 \in \PX$ such that ${\E c({X_{P_0}}) \le \Gamma}$, where ${X_{P_0}}$ denotes the random variable {subject} to $P_0$, and
\begin{align}
 \sup \left\{S \,|\, G_w(R,S|P_0) \le \varepsilon  \right\} \ge {\overline{D}_\varepsilon(R|\Gamma)} - \delta. \label{eq:dominant_prob}
\end{align}
We shall show that $S = {\overline{D}_\varepsilon(R|\Gamma)} - 4 \delta$ is {$(\varepsilon, R| \Gamma)$-achievable}. 

Fix a $P_0$ satisfying \eqref{eq:dominant_prob} and a constant $\gamma >0$ such that $\delta > 2\gamma$.
By Lemma \ref{lem:Feinstein_bound} with 
\begin{align}
\frac{1}{n} \log M_n = R + \frac{1}{\sqrt{n}} ({\overline{D}_\varepsilon(R|\Gamma)} - 4 \delta)
\end{align}
and $\eta = \frac{\gamma}{\sqrt{n}}$, we have
\begin{align}
\varepsilon_n \le \Pr \left\{ \frac{1}{n} \log \frac{W^n(Y^n|X^n)}{P_{Y^n}(Y^n)} \le R + \frac{1}{\sqrt{n}} \left ({\overline{D}_\varepsilon(R|\Gamma)} - 4 \delta  + \gamma \right) \right\}  + e^{- \sqrt{n} \gamma}. \label{eq:upper_ineq1}
\end{align}
We choose a type $P_n$ {on $\mathcal{X}^n$} {so as to be specified by \eqref{eq:cost_satisfy}--\eqref{eq:type_convergence}.}
Let $X^n$ be the uniformly distributed input random variable on $T_n$, {defined to be the set of all} sequences $\vect{x} \in \mathcal{X}^n$ of type $P_n$. 
Then, we have
\begin{align}
P_{Y_\theta^n}(\vect{y}) \le (n+1)^{|\mathcal{X}|} (P_n W_\theta)^{\times n} (\vect{y})~~{(\forall \vect{y} \in \mathcal{Y}^n)} \label{eq:PY_upper_bound}
\end{align}
by \eqref{eq:ineq2c}.
Then, by \eqref{eq:upper_ineq1}, we obtain
\begin{align}
\limsup_{n \rightarrow \infty} \varepsilon_n &\le \limsup_{n \rightarrow \infty}  \Pr \left\{ \frac{1}{n} \log \frac{W^n(Y^n|X^n)}{P_{Y^n}(Y^n)} \le R + \frac{1}{\sqrt{n}} \left ({\overline{D}_\varepsilon(R|\Gamma)} - 4 \delta  + \gamma \right) \right\}  \nonumber \\ 
&\le \limsup_{n \rightarrow \infty}  \int_{{\Theta}} dw(\theta) \Pr \left\{ \frac{1}{n} \log \frac{W^n(Y_\theta^n|X^n)}{P_{Y^n}(Y_\theta^n)} \le R + \frac{1}{\sqrt{n}} \left ({\overline{D}_\varepsilon(R|\Gamma)} - 4 \delta  + \gamma \right) \right\} \nonumber \\ 
&= \limsup_{n \rightarrow \infty} \left[  \int_{{\Theta}_n^*} dw(\theta) \Pr \left\{ \frac{1}{n} \log \frac{W^n(Y_\theta^n|X^n)}{P_{Y^n}(Y_\theta^n)} \le R + \frac{1}{\sqrt{n}} \left ({\overline{D}_\varepsilon(R|\Gamma)} - 4 \delta  + \gamma \right) \right\} \right. \nonumber \\
&~~+ \left. \int_{{\Theta} - {\Theta}_n^*} dw(\theta) \Pr \left\{ \frac{1}{n} \log \frac{W^n(Y_\theta^n|X^n)}{P_{Y^n}(Y_\theta^n)} \le R + \frac{1}{\sqrt{n}} \left ({\overline{D}_\varepsilon(R|\Gamma)} - 4 \delta  + \gamma \right) \right\} \right] \nonumber  \\ 
&\le \limsup_{n \rightarrow \infty} \int_{{\Theta}_n^*} dw(\theta) \Pr \left\{ \frac{1}{n} \log \frac{W^n(Y_\theta^n|X^n)}{P_{Y^n}(Y_\theta^n)} \le R + \frac{1}{\sqrt{n}} \left ({\overline{D}_\varepsilon(R|\Gamma)} - 4 \delta  + \gamma \right) \right\}  \nonumber \\
&~~+ \limsup_{n \rightarrow \infty}  \int_{{\Theta} - {\Theta}_n^*} dw(\theta) \Pr \left\{ \frac{1}{n} \log \frac{W^n(Y_\theta^n|X^n)}{P_{Y^n}(Y_\theta^n)} \le R + \frac{1}{\sqrt{n}} \left ({\overline{D}_\varepsilon(R|\Gamma)} - 4 \delta  + \gamma \right) \right\}  \nonumber  \\ 
&= \limsup_{n \rightarrow \infty} \int_{{\Theta}_n^*} dw(\theta) \Pr \left\{ \frac{1}{n} \log \frac{W^n(Y_\theta^n|X^n)}{P_{Y^n}(Y_\theta^n)} \le R + \frac{1}{\sqrt{n}} \left ({\overline{D}_\varepsilon(R|\Gamma)} - 4 \delta  + \gamma \right) \right\},  \label{eq:upper_ineq2}
\end{align}
where the last equality is due to \eqref{eq:non_dominant_component}.

Now by {\eqref{eq:PY_upper_bound} and} Lemma \ref{lem:mixed_upper_decomposition} with $z_n = R + \frac{1}{\sqrt{n}} \left ({\overline{D}_\varepsilon(R|\Gamma)} - 4 \delta  + \gamma \right)$,
\begin{align}
\limsup_{n \rightarrow \infty} \varepsilon_n &\le \limsup_{n \rightarrow \infty} \int_{{\Theta}_n^*} dw(\theta) \Pr \left\{ \frac{1}{n} \log \frac{W_\theta^n(Y_\theta^n|X^n)}{P_{Y_\theta^n}(Y_\theta^n)} \le R + \frac{1}{\sqrt{n}} \left ({\overline{D}_\varepsilon(R|\Gamma)} - 4 \delta  + 2 \gamma \right) + \frac{1}{\sqrt[4]{n^3}} \right\} \nonumber \\
 &\le \limsup_{n \rightarrow \infty} \int_{{\Theta}_n^*} dw(\theta) \Pr \left\{ \frac{1}{n} \log \frac{W_\theta^n(Y_\theta^n|X^n)}{(P_n W_\theta)^{\times n}(Y_\theta^n)} \le R + \frac{1}{\sqrt{n}} \left ({\overline{D}_\varepsilon(R|\Gamma)} - 4 \delta  + 2 \gamma \right) \right. \nonumber \\
&~~~~~~~~~~~~~~~~~~~~~~~~~~~~~~~\left. + \frac{1}{\sqrt[4]{n^3}} + \frac{|\mathcal{X}| \log (n+1)}{n} \right\} \nonumber \\
&\le \limsup_{n \rightarrow \infty} \int_{{\Theta}_n^*} dw(\theta) \Pr \left\{ \frac{1}{n} \log \frac{W_\theta^n(Y_\theta^n|X^n)}{(P_n W_\theta)^{\times n}(Y_\theta^n)} \le R + \frac{1}{\sqrt{n}} \left ({\overline{D}_\varepsilon(R|\Gamma)} - 3 \delta \right) \right\}. 
\label{eq:upper_ineq3}
\end{align}
Since 
\begin{align}
&\int_{{\Theta}_n^*} dw(\theta) \Pr \left\{ \frac{1}{n} \log \frac{W_\theta^n(Y_\theta^n|X^n)}{(P_n W_\theta)^{\times n}(Y_\theta^n)} \le R + \frac{1}{\sqrt{n}} \left ({\overline{D}_\varepsilon(R|\Gamma)} - 3 \delta \right) \right\} \nonumber \\
& ~~= \sum_{\vect{x} \in T_n } \Pr\{ X^n = \vect{x} \} \int_{{\Theta}_n^*} dw(\theta)  \Pr \left\{ \frac{1}{n} \log \frac{W_\theta^n(Y_\theta^n|\vect{x})}{(P_n W_\theta)^{\times n}(Y_\theta^n)} \le R + \frac{1}{\sqrt{n}} \left ({\overline{D}_\varepsilon(R|\Gamma)} - 3 \delta \right) \Big| X^n = \vect{x} \right\},  
\label{eq:upper_ineq4}
\end{align}
there exists an $\vect{x}_n \in T_n$ such that
\begin{align}
\limsup_{n \rightarrow \infty} \varepsilon_n &\le \limsup_{n \rightarrow \infty} \int_{{\Theta}_n^*} dw(\theta) \Pr \left\{ \frac{1}{n} \log \frac{W_\theta^n(Y_\theta^n| \vect{x}_n)}{(P_n W_\theta)^{\times n}(Y_\theta^n)} \le R + \frac{1}{\sqrt{n}} \left ({\overline{D}_\varepsilon(R|\Gamma)} - 3 \delta \right) \Big| X^n = \vect{x}_n \right\} \nonumber \\
& \le \limsup_{n \rightarrow \infty} \int_{{\Theta}} dw(\theta) \Pr \left\{ \frac{1}{n} \log \frac{W_\theta^n(Y_\theta^n| \vect{x}_n)}{(P_n W_\theta)^{\times n}(Y_\theta^n)} \le R + \frac{1}{\sqrt{n}} \left ({\overline{D}_\varepsilon(R|\Gamma)} - 3 \delta \right) \Big| X^n = \vect{x}_n \right\} \nonumber \\
& \le \int_{{\Theta}} dw(\theta) \limsup_{n \rightarrow \infty}  \Pr \left\{ \frac{1}{n} \log \frac{W_\theta^n(Y_\theta^n| \vect{x}_n)}{(P_n W_\theta)^{\times n}(Y_\theta^n)} \le R + \frac{1}{\sqrt{n}} \left ({\overline{D}_\varepsilon(R|\Gamma)} - 3 \delta \right) \Big| X^n = \vect{x}_n \right\},  \label{eq:upper_ineq5}
\end{align}
where the last inequality is due to Fatou's lemma.

Now, again since 
\begin{align}
{\frac{1}{n} \log \frac{W_\theta^n(Y_\theta^n| \vect{x}_n)}{(P_n W_\theta)^{\times n}(Y_\theta^n)} } \nonumber
\end{align}
 is a sum of  {conditionally} independent random variables {given $X^n = \vect{x}_n$}, {by virtue of \eqref{eq:type_convergence}}, {\eqref{eq:independent_sum}--\eqref{eq:independent_sum_variance}} {and the weak law of large numbers}, we have
\begin{align}
\limsup_{n \rightarrow \infty} \Pr \left\{ \frac{1}{n} \log  \frac{W_\theta^n(Y_\theta^n| \vect{x}_n)}{(P_n W_\theta)^{\times n}(Y_\theta^n)} \le R + \frac{1}{\sqrt{n}} \left ({\overline{D}_\varepsilon(R|\Gamma)} - 3 \delta \right)  \Big| X^n = \vect{x}_n \right\} = \left\{
\begin{array}{cc}
1, & \mathrm{if}~\theta \in {\Theta}_1 \\ 
0, & \mathrm{if}~\theta \in {\Theta}_3
\end{array}
\right. , \label{eq:wlLN}
\end{align}
where ${\Theta}_i~(i=1,2,3)$ is defined as
\begin{align}
{\Theta}_1 &:= \{ \theta \in {\Theta} | I({P_0}, W_\theta) < R \}, \label{eq:Phi_1b} \\
{\Theta}_2 &:= \{ \theta \in {\Theta} | I({P_0}, W_\theta) = R \}, \label{eq:Phi_2b} \\
{\Theta}_3 &:= \{ \theta \in {\Theta} | I({P_0}, W_\theta) > R \}. \label{eq:Phi_3b}
\end{align}
Thus,
\begin{align}
\limsup_{n \rightarrow \infty} \varepsilon_n &\le  \int_{{\Theta}_1} dw(\theta)  \nonumber \\
 &~~~+ \int_{{\Theta}_2} dw(\theta) \limsup_{n \rightarrow \infty}  \Pr \left\{ \frac{1}{n} \log \frac{W_\theta^n(Y_\theta^n| \vect{x}_n)}{(P_n W_\theta)^{\times n}(Y_\theta^n)} \le R + \frac{1}{\sqrt{n}} \left ({\overline{D}_\varepsilon(R|\Gamma)} - 3 \delta \right) \Big| X^n = \vect{x}_n \right\}. 
\label{eq:upper_ineq6}
\end{align}
Denoting the second term on the right-hand side by $B$, we have 
\begin{align}
B &= \int_{\{ \theta | I(P_0, W_\theta) = R \}} dw(\theta) \limsup_{n \rightarrow \infty}  \Pr \left\{ \frac{1}{n} \log \frac{W_\theta^n(Y_\theta^n| \vect{x}_n)}{(P_n W_\theta)^{\times n}(Y_\theta^n)} \le  I(P_0, W_\theta) + \frac{1}{\sqrt{n}} \left ({\overline{D}_\varepsilon(R|\Gamma)} - 3 \delta \right) \Big| X^n = \vect{x}_n \right\},  \nonumber \\
 &=  \int_{\{ \theta | I(P_0, W_\theta) = R \}} dw(\theta) \limsup_{n \rightarrow \infty}  \Pr \left\{ \frac{1}{\sqrt{n}} \left( \log \frac{W_\theta^n(Y_\theta^n| \vect{x}_n)}{(P_n W_\theta)^{\times n}(Y_\theta^n)}  - n  I({P_n}, W_\theta) \right) \right. \nonumber \\
 &~~~~~~~~~~~~~\left. \le {\overline{D}_\varepsilon(R|\Gamma)} - 3 \delta  + \sqrt{n} (I(P_0, W_\theta) - I(P_n, W_\theta)) \Big| X^n = \vect{x}_n \right\}.
\label{eq:upper_ineq6b}
\end{align}
{Now, we notice that, owing to \eqref{eq:type_diff},
\begin{align}
\lim_{n \rightarrow \infty} \sqrt{n} \big( I(P_0, W_\theta) - I(P_n, W_\theta) \big) =0 \label{eq:good_conv_speed}
\end{align}
and 
\begin{align}
\lim_{n \rightarrow \infty} V_{\theta, P_n} = V_{\theta, P_0},
\end{align}
and therefore}, {for $\theta \in \Theta_2$ with $V_{\theta, P_0} > 0$} the central limit theorem assures {that}
\begin{align}
\limsup_{n \rightarrow \infty}  & \Pr \left\{ \frac{1}{\sqrt{n}} \left( \log \frac{W_\theta^n(Y_\theta^n| \vect{x}_n)}{(P_n W_\theta)^{\times n}(Y_\theta^n)}  - n  I({P_n}, W_\theta) \right) \right. \nonumber \\
 &~~~~~~~~~~~~~\left. \le {\overline{D}_\varepsilon(R|\Gamma)} - 3 \delta  + \sqrt{n} (I(P_0, W_\theta) - I(P_n, W_\theta)) \Big| X^n = \vect{x}_n \right\} \nonumber \\
& = \limsup_{n \rightarrow \infty}  \Pr \left\{ \frac{1}{\sqrt{n}} \left( \log \frac{W_\theta^n(Y_\theta^n| \vect{x}_n)}{(P_n W_\theta)^{\times n}(Y_\theta^n)}  - n  I({P_n}, W_\theta) \right)  \le {\overline{D}_\varepsilon(R|\Gamma)} - 3 \delta \Big| X^n = \vect{x}_n \right\} \nonumber \\
&\, {\le \Psi_{\theta, P_0}({\overline{D}_\varepsilon(R|\Gamma)} - 2  \delta).} \label{eq:uppr_ineq6c}
\end{align}
{For $\theta \in \Theta_2$ with $V_{\theta, P_0} = 0$, we interpret $\Psi_{\theta, P_0}(z)$ as the step function which takes zero for $z < 0$ and one otherwise.
It is easily verified that \eqref{eq:uppr_ineq6c} also holds for such $\theta \in \Theta_2$,}
and hence
\begin{align}
B &\le \int_{\{ \theta | I(P_0, W_\theta) = R \}} \Psi_{\theta, P_0}({\overline{D}_\varepsilon(R|\Gamma)} - 2 \delta) \, dw(\theta) . \label{eq:upper_ineq6c2}
\end{align}
Thus, by \eqref{eq:upper_ineq6},
\begin{align}
\limsup_{n \rightarrow \infty} \varepsilon_n &\le  \int_{\{ \theta | I(P_0, W_\theta) < R \}} dw(\theta)  + \int_{\{ \theta | I(P_0, W_\theta) = R \}} \Psi_{\theta, P_0}({\overline{D}_\varepsilon(R|\Gamma)} - 2 \delta) \, dw(\theta)  \le \varepsilon,
\label{eq:upper_ineq6d}
\end{align}
where the last inequality follows from \eqref{eq:dominant_prob}, implying that ${\overline{D}_\varepsilon(R|\Gamma)} - 4 \delta$ is {$(\varepsilon, R|\Gamma)$}-achievable.
\QED

\section{Coding Theorems for Well-Ordered Mixed Memoryless Channel} \label{sect:well-ordered_channel}

\subsection{Well-Ordered Mixed Memoryless Channel} \label{sect:introduction_WO_channel}

So far, in Sect.\ \ref{sect:general_mixed_DMC_achievable_2nd}, we have established Theorem \ref{theo:general_mixed_DMC_achievable_2nd} on the second-order capacity for the mixed memoryless channel with general mixture; however, unfortunately, this theorem lacks the converse part. 
Thus, in this section, we are led to introduce a subclass of {general} mixed memoryless channels for which the second-order coding theorem is established, including both of the direct and converse parts.

\begin{e_defin} \label{def:WO_mixed_channel}
Let  $W_{\Theta} = \{ W_\theta:\mathcal{X} \rightarrow \mathcal{Y} \}_{\theta \in \Theta}$ be a family of stationary memoryless channels.
Let {${c_{\theta, \Gamma}}$ denote} the capacity of component channel $W_\theta$ {with cost constraint $\Gamma~(\ge \Gamma_0)$}, that is,
\begin{align}
{c_{\theta, \Gamma}} = \max_{P: \E c(X_P) \le \Gamma} I(P, W_\theta),
\end{align} 
and let $\Pi_{\theta, \Gamma}$ denote the set of input probability distributions {$P$} on $\mathcal{X}$ that achieve ${c_{\theta, \Gamma}} $.
It should be noted that $\Pi_{\theta, \Gamma}$ is a bounded closed set.
If $W_\Theta$ is closed and, {for any $\theta \in \Theta$ and any $P \in \Pi_{\theta, \Gamma}$, it holds that
\begin{align}
{c_{\theta, \Gamma}} &= I(P, W_{\theta'}) ~~~\mathrm{for}~~\theta' \in \Theta~~\mathrm{s.t.}~~{c_{\theta, \Gamma}} = {c_{\theta', \Gamma}}~~\mbox{and} \nonumber \\
{c_{\theta, \Gamma}} &< I(P, W_{\theta'}) ~~~\mathrm{for}~~\theta' \in \Theta~~\mathrm{s.t.}~~{c_{\theta, \Gamma}} < {c_{\theta', \Gamma}}, \label{eq:cond_WO_DMC}
\end{align}}%
then $W_\Theta$ is said to be \emph{well-ordered with cost constraint $\Gamma$}, or simply \emph{$\Gamma$-well-ordered}.
A mixed memoryless channel {$\vect{W}$} with $\Gamma$-well-ordered $W_\Theta$ is referred to as \emph{$\Gamma$-well-ordered} mixed memoryless channel.
\QED
\end{e_defin}

\smallskip
\begin{e_rema} \label{rema:optimum_distribution_WO}
For a $\Gamma$-well-ordered mixed memoryless channel, {it is not difficult to check} that
\begin{align}
\Pi_{\theta, \Gamma} = \Pi_{\theta', \Gamma}~~ \mathrm{if}~~{c_{\theta, \Gamma}} = {c_{\theta', \Gamma}}~\mathrm{for}~\theta, \theta' \in \Theta,
\end{align}
that is, two component channels with equal capacity have the same set of capacity-achieving input distributions.
\QED
\end{e_rema}

\smallskip
\begin{e_rema}
{The assumption that $W_\Theta$ is closed is made just due to a
technical reason. Even in the case where $W_\Theta$ is not closed,
if its closure denoted by $W_{\overline{\Theta}}$ (with extended
parameter space $\overline{\Theta}$) is $\Gamma$-well-ordered,
all coding theorems we shall establish also hold for the mixed channel $\vect{W}$ with the original
$W_\Theta$.}%
\QED
\end{e_rema}

\smallskip
\begin{e_exam} 
For two channels $W_\theta$ and $W_{\theta'}$, channel $W_{\theta'}$ is said to be \emph{more capable} than $W_\theta$ if $I(P, W_\theta) \le I(P, W_{\theta'})$ for all $P \in \PX$ \cite{Csiszar-Korner2011}.
If $W_{\theta'}$ is more capable than $W_\theta$ for all $\theta, \theta' \in \Theta$ such that {${c_{\theta}} \le {c_{\theta'}}$, then $W_\Theta$ is $\Gamma$-well-ordered  for all $\Gamma \ge {\Gamma_0}$}, where ${c_{\theta}}$ denotes the capacity of $W_\theta$ with no cost constraints. The followings are examples of such $W_\Theta$:
\begin{itemize}
\item A family of binary symmetric channels which forms a closed set. 
\item More generally, a closed set of additive noise channels for which additive noise $Z \sim {W_\theta(\cdot|\cdot)}$ is a degraded version of  additive noise $Z' \sim {W_{\theta'} (\cdot|\cdot)}$ for all $\theta, \theta' \in \Theta$ such that ${c_{\theta}} \le {c_{\theta'}}$.
\QED
\end{itemize}
\end{e_exam}

\smallskip
\begin{e_exam} 
In the special case of $\Gamma = + \infty$ (that is, without cost constraints), we may find much more examples of $\Gamma$-well-ordered $W_\Theta$. 
A family of output-symmetric channels which forms a closed set is $\Gamma$-well-ordered since the capacity-achieving input distribution is uniform on $\mathcal{X}$ and unique (cf.\ Shannon \cite{Shannon48}). 
\QED
\end{e_exam}

\smallskip
{Set} $E_{\Theta, \Gamma} := \{ {c_{\theta, \Gamma}} \, | \, \theta \in \Theta \}$.
We show an important property of $\Gamma$-well-ordered mixed memoryless channels.
\begin{e_lem} \label{lem:E_closed}
If $W_\Theta$ is closed, then $E_{\Theta, \Gamma}$ is bounded and closed for all $\Gamma \ge {\Gamma_0}$.
\end{e_lem}
(Proof)~~Boundedness of $E_{\Theta, \Gamma}$ is {obvious}, so we shall show its closedness.
Let a function $f: \PXY \rightarrow [0, +\infty)$ be defined as
\begin{align}
f(W) := \max_{P \in \mathcal{P}_C} I(P, W)
\end{align}
for a given closed convex set $ \mathcal{P}_C \subseteq \PX $, where  $\PXY$ denotes the set of all {channel matrices $W: \mathcal{X} \rightarrow \mathcal{Y}$}.
Since $I(P,W)$ is continuous with respect to $(P, W)$, the $f(W)$ is a continuous function of $W$.
The image of a closed set by a continuous function is also closed. {Hence,} since $ \mathcal{P}_C := \{ P \in \PX | \, \E c(X_P) \le \Gamma\}$ is closed and convex, we can conclude that $E_{\Theta, \Gamma} = f(W_\Theta)$ is closed.   
\QED

\subsection{Coding Theorems} \label{sect:coding_theorem_WO_channel}

We first provide a characterization of the first-order capacity $C_\varepsilon(\Gamma)$, which is different from the one in Theorem \ref{theo:general_mixed_DMC_e-cap}, for $\Gamma$-well-ordered mixed memoryless channels.  
This alternative characterization is of simpler form and is of great use to analyze the second-order capacity later.  

\smallskip
\begin{e_theo}\label{theo:WO_mixed_DMC_e-cap}
Let $\vect{W}$ be a $\Gamma$-well-ordered mixed memoryless channel with {general} measure $w$.
For any fixed $\varepsilon \in [0,1)$ {and $\Gamma \ge {\Gamma_0}$}, the {first-order} {$(\varepsilon|\Gamma)$}-capacity is given by
\begin{align}
{C_\varepsilon(\Gamma)} =  \sup \left\{ R \, \Big|  \int_{\{ \theta|\, {c_{\theta, \Gamma}} < R \}} dw(\theta) \le \varepsilon  \right\}. \label{eq:e-Cap_WO_general_mixture}
\end{align}
\QED
\end{e_theo}

\smallskip
\begin{e_rema}
Due to the closedness of $E_{\Theta, \Gamma}$, for every $\varepsilon \in [0, 1)$ there exists some $\bartheta \in \Theta$ such that $C_\varepsilon(\Gamma) = {c_{\bartheta, \Gamma}}$.
This fact is shown in the the proof of the converse part of Theorem \ref{theo:WO_mixed_DMC_e-cap} in Sect.\ \ref{sect:proof_WO_mixed_DMC_e-cap}.
\QED
\end{e_rema}
 

\smallskip
\begin{e_rema}
The characterization \eqref{eq:e-Cap_WO_general_mixture} with $\Gamma = +\infty$ is a generalization of  the one given by Winkelbauer \cite{Winkelbauer71a} in the sense that the class of $\Gamma$-well-ordered mixed channels with $\Gamma = +\infty$ is wider than the class of \emph{regular decomposable} channels with stationary memoryless components.
On the other hand, the regular decomposability allows component channels to be stationary and ergodic, which means that the characterization \eqref{eq:e-Cap_WO_general_mixture} with  $\Gamma = +\infty$ is a particularization of the one given in \cite{Winkelbauer71a}.
\QED
\end{e_rema}%

\smallskip
Now, we turn to discussing the second-order capacity {of} $\Gamma$-well-ordered mixed memoryless channels.
{In contrast} to {mixed memoryless channels with general mixture}, for which only the direct part of the second-order coding theorem (Theorem \ref{theo:general_mixed_DMC_achievable_2nd}) has been given, $\Gamma$-well-orderedness allows us to establish the converse theorem as well. 
\smallskip
\begin{e_theo} \label{theo:WO_mixed_DMC_2nd}
Let $\vect{W}$ be a $\Gamma$-well-ordered mixed memoryless channel with {general} measure $w$.
Then, for $\varepsilon \in [0,1)$, $\Gamma \ge {\Gamma_0}$, and $R \ge 0$, 
it holds that
\begin{align}
D_\varepsilon(R|\Gamma)  &=  \sup_{ P : \E c(X_P) \le \Gamma } \sup \left\{ S \, \Big| \, \int_{\{\theta | I(P, W_\theta) < R \}} d w(\theta) +  \int_{\{\theta | I(P, W_\theta) =  R \}}\Psi_{\theta, P}(S) d w(\theta) \le  \varepsilon  \right\} \nonumber \\
 &=  \sup_{ P \in \Pi_{\bartheta, \Gamma} } \sup \left\{ S \, \Big| \, \int_{\{\theta | I(P, W_\theta) < R \}} d w(\theta) +  \int_{\{\theta | I(P, W_\theta) =  R \}}\Psi_{\theta, P}(S) d w(\theta) \le  \varepsilon  \right\} \nonumber \\
 &=  \sup_{ P \in \Pi_{\bartheta, \Gamma} } \sup \left\{ S \, \Big| \, \int_{\{\theta | {c_{\theta, \Gamma}}< R \}} d w(\theta) +  \int_{\{\theta | {c_{\theta, \Gamma}} =  R \}}\Psi_{\theta, P}(S) d w(\theta) \le  \varepsilon  \right\}, \label{eq:WO_2nd_order_general_mixture}
\end{align}
where $\bartheta \in \Theta$ gives the $(\varepsilon|\Gamma)$-capacity, that is $C_\varepsilon(\Gamma) = {c_{\bartheta, \Gamma}}$.
\QED
\end{e_theo}

\smallskip
\begin{e_rema}
Formula \eqref{eq:WO_2nd_order_general_mixture} has been established for the case of $|\Theta| < + \infty$ by Yagi and Nomura \cite{Yagi-Nomura2014b}. When the component channels are output-symmetric and $\Gamma = + \infty$, the first supremum (with respect to $P$) on the right-hand side of \eqref{eq:WO_2nd_order_general_mixture} is attained by only the \emph{uniform} inputs, which may facilitate the proof of the coding theorem.
\QED
\end{e_rema}

\smallskip
\begin{e_rema} \label{rema:canonical_rep_WO}
It is not difficult to check from formula \eqref{eq:e-Cap_WO_general_mixture}  that
\begin{align}
{\int_{\{ \theta | {c_{\theta, \Gamma}} < C_\varepsilon (\Gamma)\}} dw(\theta)} \,  & \, {\le  \varepsilon,} \label{eq:pre_majorization2} \\ 
\int_{\{ \theta | {c_{\theta, \Gamma}} \le C_\varepsilon (\Gamma)\}} dw(\theta) &\ge  \varepsilon \label{eq:post_majorization2}
\end{align}
hold, and like in Remark \ref{rema:canonical_rep} here also we may consider the following canonical equation for $S$:
\begin{align}
\int_{\Theta} dw(\theta) \lim_{n \rightarrow \infty} \Psi_{\theta, P} \big( \sqrt{n} (C_\varepsilon(\Gamma) - {c_{\theta, \Gamma}}) + S \big) = \varepsilon. \label{eq:canonical_equation}
\end{align}
Notice here, in view of \eqref{eq:pre_majorization2} and \eqref{eq:post_majorization2}, that equation \eqref{eq:canonical_equation} always has a solution. 
Let $S_P(\varepsilon)$ denote the solution of this equation, {where}
$S_P(\varepsilon) = + \infty$ if the solution is not unique.
Then, the $D_\varepsilon \big(C_\varepsilon(\Gamma)| \Gamma\big)$ (i.e., $R = C_\varepsilon(\Gamma)$) can be rewritten in a simpler form as 
\begin{align}
D_\varepsilon \big(C_\varepsilon(\Gamma) | \Gamma\big) = \sup_{ P \in \Pi_{\bartheta, \Gamma} }  S_P(\varepsilon),
\end{align}
which is again sometimes preferable to the expression in \eqref{eq:WO_2nd_order_general_mixture}.%
\QED
\end{e_rema}

\subsection{Proof of Theorem \ref{theo:WO_mixed_DMC_e-cap}} \label{sect:proof_WO_mixed_DMC_e-cap}

\noindent
(Proof of Converse Part)

By definition, it holds that $I(P, W_\theta) \le {c_{\theta, \Gamma}}$ for all $\theta \in \Theta$ if $P$ satisfies $\E c(X_P) \le \Gamma$.
Therefore, by \eqref{eq:e-Cap_general_mixture} in Theorem \ref{theo:general_mixed_DMC_e-cap}, we have
\begin{align}
{C_\varepsilon(\Gamma)} &\le  \sup_{ {P} : {\E c({X_P}) \le \Gamma}} \sup \left\{ R \, \Big|  \int_{\{ \theta|\, {c_{\theta, \Gamma}} < R \}} dw(\theta) \le \varepsilon  \right\} \nonumber \\
&=  \sup \left\{ R \, \Big|  \int_{\{ \theta|\, {c_{\theta, \Gamma}} < R \}} dw(\theta) \le \varepsilon  \right\}. 
\end{align}
\QED

\medskip
\noindent
(Proof of Direct Part)

Set
\begin{align}
\overline{R} &=\sup \left\{ R \, \Big|  \int_{\{ \theta|\, {c_{\theta, \Gamma}} < R \}} dw(\theta) \le \varepsilon  \right\} 
\end{align}
for notational simplicity.
Consider an increasing sequence $R_1 \le R_2 \le \cdots \rightarrow \overline{R}$ such that
\begin{align}
 \int_{\{ \theta|\, {c_{\theta, \Gamma}} < R_i \}} dw(\theta) \le \varepsilon~~(\forall i = 1, 2, \cdots). 
\end{align}
Then, we have 
\begin{align}
 \int_{\{ \theta|\, {c_{\theta, \Gamma}} < \overline{R} \}} dw(\theta) \le \varepsilon 
\end{align}
by the continuity of probability measures.
Now suppose that $\overline{R}$ is not an accumulation point of $E_{\Theta, \Gamma}$ {to show} a contradiction.
Then, there exists some $\nu > 0$ such that
\begin{align}
(\overline{R} - \nu, \overline{R} + \nu) \cap E_{\Theta, \Gamma} = \emptyset.
\end{align}
{
This implies that $\{\theta |\, \overline{R} \le  {c_{\theta, \Gamma}} < \overline{R} + \nu \} = \emptyset$, and hence, we have
\begin{align}
 \int_{\{ \theta|\, {c_{\theta, \Gamma}} < \overline{R} + \nu \}} dw(\theta) &=  \int_{\{ \theta|\, {c_{\theta, \Gamma}} < \overline{R} \}} dw(\theta)  \le \varepsilon, 
\end{align}}
 which contradicts the definition of $\overline{R}$.
 Therefore, $\overline{R}$ is an accumulation point of  $E_{\Theta, \Gamma}$.
 Since $E_{\Theta, \Gamma}$ is a closed set by Lemma \ref{lem:E_closed}, it holds that $\overline{R} \in E_{\Theta, \Gamma}$, and there exists some $\bartheta \in \Theta$ such that $\overline{R} = {c_{\bartheta, \Gamma}}$.
 
Fixing $P  \in \Pi_{\bartheta, \Gamma}$ arbitrarily, we have
\begin{align}
& \int_{\{ \theta|\, I(P, W_\theta) < \overline{R} \}} dw(\theta) \nonumber \\
&~~= \int_{\{ \theta|\, I(P, W_\theta) < \overline{R}, \, {c_{\theta, \Gamma}} < \overline{R} \}} dw(\theta) + \int_{\{ \theta|\, I(P, W_\theta) < \overline{R}, \, {c_{\theta, \Gamma}} \ge \overline{R} \}} dw(\theta) \nonumber \\
&~~= \int_{\{ \theta|\, I(P, W_\theta) < \overline{R}, \, {c_{\theta, \Gamma}} <  \overline{R} \}} dw(\theta), \label{eq:eq1}
\end{align}
where the last equality follows from the fact that there are no $\theta \in \Theta$ such that
${c_{\theta, \Gamma}} \ge \overline{R} =  {c_{\bartheta, \Gamma}}$ and $I(P, W_\theta) < \overline{R}$ for $P \in \Pi_{\bartheta, \Gamma}$ by the definition of $\Gamma$-well-orderedness.
Noticing that $\{ \theta | \, I(P, W_\theta) < \overline{R}, {c_{\theta, \Gamma}} < \overline{R} \} = \{ \theta | \, {c_{\theta, \Gamma}} < \overline{R} \}$ for $P \in \Pi_{\bartheta, \Gamma}$ in \eqref{eq:eq1}, we have
\begin{align}
\int_{\{ \theta|\, I(P, W_\theta) < \overline{R} \}} dw(\theta) = \int_{\{ \theta|\, {c_{\theta, \Gamma}} < \overline{R} \}} dw(\theta) \le \varepsilon, 
\end{align}
and formula \eqref{eq:e-Cap_general_mixture} in Theorem \ref{theo:general_mixed_DMC_e-cap} indicates that $\overline{R} \le C_{\varepsilon}(\Gamma)$.
\QED

\subsection{Proof of Theorem \ref{theo:WO_mixed_DMC_2nd}}  \label{sect:proof_WO_mixed_DMC_2nd}

\noindent
(Proof of Direct Part)

It apparently holds, with $G_w(R,S|P)$ as in \eqref{eq:func_Gw}, that
\begin{align}
&\sup_{ P : \E c(X_P) \le \Gamma } \sup \left\{ S \, \Big| \, G_w(R, S|P)  \le  \varepsilon  \right\}  \ge \sup_{ P \in \Pi_{\bartheta, \Gamma} } \sup \left\{ S \, \Big| \, G_w(R, S|P)  \le  \varepsilon  \right\} \label{eq:equrivalent_expression}
\end{align}
since any $P \in \Pi_{\bartheta, \Gamma}$ satisfies cost constraint: $\E c(X_P) \le \Gamma$.
Therefore, by Theorem \ref{theo:general_mixed_DMC_achievable_2nd}, any $S$ such that
\begin{align}
S < \sup_{ P \in \Pi_{\bartheta, \Gamma} } \sup \left\{ S \, \Big| \, G_w(R, S|P)  \le  \varepsilon  \right\} \label{eq:equrivalent_expression2}
\end{align}
is $(\varepsilon, R|\Gamma)$-achievable.
\QED

\medskip
\noindent
(Proof of Converse Part)

{Although the converse part can be established on the basis of Lemmas \ref{lem:HN_bound} and \ref{lem:mixed_lower_decomposition} in a manner similar to the converse proof of Theorem \ref{theo:general_mixed_DMC_e-cap}, here instead of these lemmas,}
we use the following {simple but powerful} lower bound on the probability of decoding error, which {is of independent interest and} facilitates the proof of this converse part.

\smallskip
\begin{e_lem} \label{lem:mixed_lower_bound2}
Let $\{ Q_\theta^n\}_{\theta \in \Theta}$ be {a family of} arbitrarily fixed output distributions on $\mathcal{Y}^n$.
 Every $(n,M_n,\varepsilon_n)$ code $\Code$ for the  mixed channel $W^n$ given in \eqref{eq:mixed_channel_general_measure} satisfies
\begin{align}
\varepsilon_n  \ge \int_{\Theta}  dw(\theta) \Pr \left\{ \frac{1}{n} \log \frac{W_\theta^n(Y_\theta^n|X^n) }{Q_\theta^n(Y_\theta^n)} \le  \frac{1}{n}  \log M_n - {\eta} \right\} - e^{-n {\eta}} \label{eq:new_LB1}
\end{align}
with an arbitrary number ${\eta} > 0$, where {$X^n$ is uniformly distributed} on $\Code$.  
\end{e_lem}
(Proof)~See Appendix \ref{append:proof_of_new_lower_bound}.
\QED
\medskip
\begin{e_rema}
 {It should be noted that Lemma \ref{lem:mixed_lower_bound2} holds for arbitrary alphabets $\mathcal{X}, \mathcal{Y}$ (not necessarily finite)}.
\QED
\end{e_rema}

Since formula \eqref{eq:WO_2nd_order_general_mixture} trivially holds in the cases $R < C_{\varepsilon}(\Gamma)$ {(with $D_\varepsilon(R|\Gamma) = + \infty$)} and $R > C_{\varepsilon}(\Gamma)$ {(with $D_\varepsilon(R|\Gamma) = - \infty$)}, hereafter we shall prove only for the case $R = C_\varepsilon(\Gamma)$, which is of our main interest.
Assume that $S$ is $(\varepsilon , R | \Gamma)$-achievable. 
Then, by definition, for any given $\gamma > 0$ there exists an $(n, M_n, \varepsilon_n)$ code with cost constraint $\Gamma$ such that
\begin{align}
\frac{1}{n} \log M_n \ge R + \frac{S - \gamma}{\sqrt{n}}~~(\forall n \ge n_0 ). \label{eq:S_achievable}
\end{align}
Following a technique developed by Hayashi \cite{Hayashi2009}, let $Q_\theta^n$ be the output distribution on $\mathcal{Y}^n$ indexed by $\theta \in {\Theta}$ such that
\begin{align}
Q_\theta^n (\vect{y}) =  \sum_{P_n \in \mathcal{T}_n} \frac{(P_nW_\theta)^{\times n}(\vect{y})}{N_n + 1}+ \frac{ (P_\theta W_\theta)^{\times n}(\vect{y})}{N_n + 1}  ~~(\forall \theta \in {\Theta}, \forall \vect{y} \in \mathcal{Y}^n),  \label{eq:Hayashi_output_dist}
\end{align}
where $\mathcal{T}_n $ with $N_n = |\mathcal{T}_n|$ denotes the set of {all} types on $\mathcal{X}^n$ and $P_\theta$ is an arbitrary input distribution in $\Pi_{\theta,\Gamma}$.
It should be noted that the capacity-achieving output distribution $P_\theta W_\theta$ for $W_\theta$ is the same for all $P_\theta \in \Pi_{\theta,\Gamma}$, and this enables us to choose a particular $P_\theta \in \Pi_{\theta, \Gamma}$ later. 
Using this $\{Q_\theta^n\}_{\theta \in {\Theta}}$, we define ${Q^n}$ as in \eqref{eq:output_dist2}.
Lemma \ref{lem:mixed_lower_bound2} by replacing $\eta$ with $\frac{\gamma}{\sqrt{n}}$ assures that the sequence of $(n, M_n, \varepsilon_n)$ codes $\Code$ (satisfying cost constraint $\Gamma$) such that
\begin{align}
\varepsilon_n & \ge \int_{{\Theta}} dw(\theta) \Pr \left\{ \frac{1}{n} \log \frac{W_\theta^n(Y_\theta^n|X^n) }{Q_\theta^n(Y_\theta^n)}  \le  R + \frac{S - 2\gamma}{\sqrt{n}}   \right\} - e^{-\sqrt{n}  \gamma} \nonumber \\
& =  \sum_{\vect{x}_n \in \mathcal{X}^n} P_{X^n}(\vect{x}_n) \int_{\Theta} dw(\theta)  \Pr \left\{ \frac{1}{n} \log \frac{W_\theta^n(Y_\theta^n|\vect{x}_n) }{Q_\theta^n(Y_\theta^n)}  \le  R + \frac{S - 2\gamma}{\sqrt{n}} \Big| X^n = \vect{x}_n \right\} - e^{-\sqrt{n} \gamma}, \label{eq:new_LB2}
\end{align}
where $P_{X^n}$ is the uniform distribution on $\Code$.
This implies that there exists a codeword $\vect{x}_n$ such that
\begin{align}
\varepsilon_n & \ge  \int_{\Theta} dw(\theta) \Pr \left\{ \frac{1}{n} \log \frac{W_\theta^n(Y_\theta^n|\vect{x}_n) }{Q_\theta^n(Y_\theta^n)}  \le  R + \frac{S - 2\gamma}{\sqrt{n}} \Big| X^n = \vect{x}_n \right\} - e^{-\sqrt{n} \gamma}. \label{eq:new_LB2b}
\end{align}
Now, we partition the parameter space $\Theta$ as follows:
\begin{align}
\Theta_1 &:= \{ \theta \in \Theta \, | \, {c_{\theta, \Gamma}} < R  \}, \label{eq:WO_Phi_1} \\
\Theta_2 &:= \{ \theta \in \Theta \, | \, {c_{\theta, \Gamma}}  = R \}, \label{eq:WO_Phi_2} \\
\Theta_3 &:= \{ \theta \in \Theta \, | \, {c_{\theta, \Gamma}}  > R \}. \label{eq:WO_Phi_3}
\end{align}
Using these partitioned spaces, we further bound \eqref{eq:new_LB2b} as
\begin{align}
\varepsilon_n & \ge  \int_{\Theta_1} dw(\theta)  B_{\theta, n}
+ \int_{\Theta_2} dw(\theta)  B_{\theta, n} - e^{-\sqrt{n} \gamma}, \label{eq:new_LB3}
\end{align}
where we have set
\begin{align}
 B_{\theta, n} := \Pr \left\{ \frac{1}{n} \log \frac{W_\theta^n(Y_\theta^n|\vect{x}_n) }{Q_\theta^n(Y_\theta^n)}  \le  R + \frac{S - 2\gamma}{\sqrt{n}} \Big| X^n = \vect{x}_n \right\}. 
\end{align}
Let $P_n \in \mathcal{T}_n$ denote the type of $\vect{x}_n$ ({obviously,} this $P_n$ satisfies $\E c(X_{P_n}) \le \Gamma$, where $X_{P_n}$ denotes the random variable subject to $P_n$).
By \eqref{eq:Hayashi_output_dist}, the probability term $B_{\theta, n}$ is lower bounded in two ways as
 \begin{align}
 B_{\theta, n} \ge \Pr \left\{ \frac{1}{n} \log \frac{W_\theta^n(Y_\theta^n|\vect{x}_n) }{(P_\theta W_\theta)^{\times n}(Y_\theta^n)}  + \frac{\log (N_n + 1)}{n}\le  R + \frac{S - 2\gamma}{\sqrt{n}} \Big| X^n = \vect{x}_n \right\} =: \alpha_{\theta, n}. \label{eq:LB_alpha}
\end{align}
and
 \begin{align}
 B_{\theta, n} \ge \Pr \left\{ \frac{1}{n} \log \frac{W_\theta^n(Y_\theta^n|\vect{x}_n) }{(P_nW_\theta)^{\times n}(Y_\theta^n)}  + \frac{\log (N_n + 1)}{n}\le  R + \frac{S - 2\gamma}{\sqrt{n}} \Big| X^n = \vect{x}_n \right\} =: \beta_{\theta, n} \label{eq:LB_beta}
\end{align}
It should be noted that both $B_{\theta, n}$ and $\alpha_{\theta, n}$ do not depend on the choice of $P_\theta \in \Pi_{\theta,\Gamma}$ in \eqref{eq:Hayashi_output_dist} since $P_\theta W_\theta$ is unique.
Notice that $ \frac{1}{n} \log \frac{W_\theta^n(Y_\theta^n|\vect{x}_n)}{ (P_nW_\theta)^{\times n} (Y_\theta^n)} $ in \eqref{eq:LB_beta} (cf.\ \eqref{eq:independent_sum}) is a sum of {conditionally} independent random variables given $X^n = \vect{x}_n$ (under $W_\theta^n(\cdot|\vect{x}_n)$) with  mean $I(P_n, W_\theta)$ and variance  $V_{\theta, P_n}$, which is given as in \eqref{eq:independent_sum_variance}.
{Moreover}, 
\begin{align}
\frac{1}{n} \log \frac{W_\theta^n(Y_\theta^n|\vect{x}_n)}{ (P_\theta W_\theta)^{\times n} (Y_\theta^n)} = \frac{1}{n} \sum_{i=1}^n \log \frac{W_\theta(Y_{\theta, i} | x_i)}{ P_\theta W_\theta (Y_{\theta, i})}  \label{eq:independent_sum2}
\end{align}
is a sum of {conditionally} independent random variables given $X^n = \vect{x}_n = (x_1, x_2, \ldots, x_n)$ (under $W_\theta^n(\cdot|\vect{x}_n)$) with mean
\begin{align}
\E \left\{ \frac{1}{n} \sum_{i=1}^n \log \frac{W_\theta(Y_{\theta, i} | x_i)}{ P_\theta W_\theta (Y_{\theta, i})}  {\Big| \, X^n = \vect{x}_n } \right\} = \sum_{x \in \mathcal{X}} P_n (x) D(W_\theta(\cdot | x) \| P_\theta W_\theta) \label{eq:independent_sum_mean2}
\end{align}
and variance 
\begin{align}
 \V \left\{ \frac{1}{n} \sum_{i=1}^n \log \frac{W_\theta(Y_{\theta, i}|x_i)}{P_\theta W_\theta(Y_{\theta,i})}   {\Big| \, X^n = \vect{x}_n } \right\} & = \sum_{x \in \mathcal{X}} \sum_{y \in \mathcal{Y}} P_n (x) W_\theta (y|x) \left( \log \frac{W_\theta(y|x)}{P_ \theta W_\theta(y)} - D(W_\theta (\cdot | x) || P_\theta W_\theta) \right)^2 \nonumber \\
 &=:  V_{\theta, P_\theta}(\vect{x}_n). \label{eq:independent_sum_variance2}
\end{align}

Since $\{P_n\}_{n=1}^\infty$ is a sequence in $\PX$, which is compact, it always contains a converging subsequence $\{P_{n_1}, P_{n_2}, \cdots \}$, where $n_1 < n_2< \cdots < \rightarrow \infty$. We denote the convergent point by $P_0$;
\begin{align}
\lim_{i \rightarrow \infty} P_{n_i} = P_0,
\end{align}
where it should be noticed that $P_0$ also satisfies cost constraint: $\E c(X_{P_0}) \le \Gamma$.
For notational simplicity, we relabel $n_k$ as $m = n_1, n_2,\cdots$.
For this subsequence, we shall evaluate
\begin{align}
A^{(1)}_m := \int_{\Theta_1} dw(\theta)  B_{\theta, \nm} ~~\mathrm{and}~~A^{(2)}_m := \int_{\Theta_2} dw(\theta)  B_{\theta, \nm} ~~~(m= n_1, n_2, \cdots) \label{eq:func_A}
\end{align}
where \eqref{eq:new_LB3} is now expressed as $\varepsilon_m \ge A^{(1)}_m + A^{(2)}_m - e^{-\sqrt{m} \gamma}$.


We first evaluate $A^{(1)}_m$.
Fix $\theta \in \Theta_1$ arbitrarily. In this case, $I(P_\nm, W_\theta) \le {c_{\theta, \Gamma}} < R$, so  $\beta_{\theta, \nm}$ on the right-hand side of \eqref{eq:LB_beta} {becomes} 
  \begin{align}
 \beta_{\theta, \nm} &= \Pr \left\{ \frac{1}{\nm} \log \frac{W_\theta^\nm(Y_\theta^\nm|\vect{x}_\nm) }{(P_\nm W_\theta)^{\times \nm}(Y_\theta^\nm)}  - I(P_\nm, W_\theta) \le  R  - I(P_\nm, W_\theta) + \frac{S - 2\gamma}{\sqrt{\nm}} -  \frac{\log (N_\nm + 1)}{\nm} \Big| X^\nm = \vect{x}_\nm \right\} \nonumber \\
 & \rightarrow 1~~~ (\nm \rightarrow \infty), \label{eq:new_LB6}
\end{align}
where the convergence is due to the weak law of large numbers.
By \eqref{eq:LB_beta}, \eqref{eq:new_LB6}, and Fatou's lemma, we obtain
\begin{align}
\liminf_{\nm \rightarrow \infty} A^{(1)}_m &\ge \liminf_{\nm \rightarrow \infty} \int_{\Theta_1} dw(\theta)  \beta_{\theta, \nm}  \nonumber \\
&\ge \int_{\Theta_1} dw(\theta) \liminf_{\nm \rightarrow \infty}  \beta_{\theta, \nm} \nonumber\\
& =  \int_{\Theta_1} dw(\theta) =  \int_{\{\theta \,|\, {c_{\theta, \Gamma}} < R\}} dw(\theta). \label{eq:beta_expansion1a}
\end{align}

Next, we turn to evaluating $A^{(2)}_m$.
We consider two cases according to whether the convergent point $\displaystyle P_0 = \lim_{\nm \rightarrow \infty} P_\nm$ is in $\Pi_{\bartheta, \Gamma}$ or not, {where ${c_{\bartheta, \Gamma}} = R = C_\varepsilon(\Gamma)$.} 
More precisely, we will bound $A^{(2)}_m$ from below {in two ways} as
\begin{align}
A^{(2)}_m = \int_{\Theta_2} dw(\theta)  B_{\theta, \nm} \ge \left\{
\begin{array}{ll}
 \int_{\Theta_2} dw(\theta) \alpha_{\theta, \nm} & \mathrm{if}~ P_0 \in \Pi_{\bartheta, \Gamma} \\
 \int_{\Theta_2} dw(\theta) \beta_{\theta, \nm} & \mathrm{if}~P_0 \not\in \Pi_{\bartheta, \Gamma} 
\end{array}
\right. . \label{eq:B_LB}
\end{align}

\begin{itemize}
\item[(i)] Consider the case {of} $P_0 \not\in \Pi_{\bartheta, \Gamma}$.
{We define 
\begin{align}
\mathcal{V}_\tau &:= \{ P \,|\, I(P, W_{{\theta}}) > {c_{\bartheta, \Gamma}} - \tau \}~~(\forall \tau > 0), 
\end{align}
where ${c_{\theta, \Gamma}}= {c_{\bartheta, \Gamma}} = R = C_\varepsilon(\Gamma)$ as we are now considering the case of $\theta \in \Theta_2$.} 
Then, for each $\theta \in \Theta_2$ there exists some $\tau_\theta > 0$ such that $P_0 \not\in \mathcal{V}_{2\tau_\theta}$. This implies that $P_m \not\in \mathcal{V}_{\tau_\theta}$ for all $m > m_0$ with some positive number $m_0 > 0$.
{Then, by Chebyshev's inequality, it holds that
\begin{align}
\beta_{\theta, \nm} \ge 1- \frac{\max_{P} V_{\theta, P}}{\left( S- 2\gamma + \sqrt{\nm} \tau_\theta - \frac{\log (N_\nm + 1)}{\sqrt{\nm}} \right)^2}  ~~~~(\forall m > m_0), \label{eq:beta_LB1}
\end{align}
where {\eqref{eq:bounded_variance} holds}, indicating that $\beta_{\theta, \nm} \rightarrow 1~(\nm \rightarrow \infty)$}. 
By Fatou's lemma and \eqref{eq:LB_beta}, we obtain 
\begin{align}
\liminf_{m \rightarrow \infty } A^{(2)}_m  &\ge   \int_{\Theta_2} dw(\theta)   \liminf_{m \rightarrow \infty } B_{\theta, \nm} \nonumber \\
&\ge    \int_{\Theta_2} dw(\theta)   \liminf_{m \rightarrow \infty } \beta_{\theta, \nm} \nonumber \\
&\ge  \int_{\Theta_2} dw(\theta)   \liminf_{m \rightarrow \infty }  \left(1- \frac{\max_{P} V_{\theta, P}}{\left( S- 2\gamma + \sqrt{\nm} \tau_\theta - \frac{\log (N_\nm + 1)}{\sqrt{\nm}} \right)^2} \right) \nonumber \\
&=  \int_{\Theta_2} dw(\theta). \label{eq:beta_LB2}
\end{align}

\item[(ii)] Next, consider the case {of} $P_0 \in \Pi_{\bartheta, \Gamma}$.
{Since ${c_{\theta, \Gamma}}= {c_{\bartheta, \Gamma}}$ for $\theta \in \Theta_2$ and hence $\Pi_{\theta, \Gamma} = \Pi_{\bartheta, \Gamma}$ (cf.\ Remark \ref{rema:optimum_distribution_WO}), in \eqref{eq:Hayashi_output_dist} we can choose $P_\theta \in \Pi_{\theta, \Gamma}$ for each $\theta \in \Theta_2$ so that 
\begin{align}
\lim_{\nm \rightarrow \infty} P_m =  P_0 = P_\theta,  \label{eq:accumulation_dist}
\end{align} 
where we notice that $B_{\theta, n}$ and $\alpha_{\theta, n}$ do not depend on the choice of $P_\theta \in \Pi_{\theta,\Gamma}= \Pi_{\bartheta,\Gamma}$.}
Since {again $P_\theta \in \Pi_{\theta, \Gamma}$ and} ${c_{\theta, \Gamma}} = R = C_\varepsilon(\Gamma)$ for $\theta \in \Theta_2$,  we have
\begin{align}
\sum_{x \in \mathcal{X}} P_\nm(x) D(W_\theta (\cdot | x) \| P_\theta W_\theta) \le {c_{\theta, \Gamma}} = R \label{eq:Kuhn-Tucker}
\end{align}
by the Kuhn-Tucker theorem. Indeed, the Kuhn-Tucker theorem asserts that
for finite $\mathcal{X}$ and $\mathcal{Y}$, it holds for all $x \in \mathcal{X}$ that
\begin{align}
D(W_\theta (\cdot | x) \| P_\theta W_\theta) \le {c_{\theta, \Gamma}} + \lambda_0( c(x) - \Gamma) \label{eq:Kuhn-Tucker2}
 \end{align}
with some $\lambda_0\ge 0$ (cf.\ \cite[Lemma 3.7.1]{Han2003}).
By taking the average with $P_\nm$ for both sides of \eqref{eq:Kuhn-Tucker2}, we obtain
\begin{align}
\sum_{x \in \mathcal{X}} P_\nm(x) D(W_\theta (\cdot | x) \| P_\theta W_\theta) &\le {c_{\theta, \Gamma}} + \lambda_0 \Big( \sum_{x \in \mathcal{X}} P_\nm(x) c(x) - \Gamma \Big), \label{eq:Kuhn-Tucker3}
 \end{align}
 which implies the inequality in \eqref{eq:Kuhn-Tucker}  since $P_\nm$ satisfies cost constraint: $\E c(X_{P_\nm}) \le \Gamma$.
 By \eqref{eq:Kuhn-Tucker}, we have
\begin{align}
\alpha_{\theta, \nm} &= \Pr \left\{ \frac{1}{\sqrt{\nm}} \left( \log \frac{W_\theta^\nm(Y_\theta^\nm|\vect{x}_\nm) }{(P_\theta W_\theta)^{\times \nm}(Y_\theta^\nm)}  - \nm R \right) \le  S - 2\gamma -   \frac{\log (N_\nm + 1)}{\sqrt{\nm}}\Big| X^\nm = \vect{x}_\nm \right\} \nonumber \\
& \ge  \Pr \left\{ \frac{1}{\sqrt{\nm}} \left( \log \frac{W_\theta^\nm(Y_\theta^\nm|\vect{x}_\nm) }{(P_\theta W_\theta)^{\times \nm}(Y_\theta^\nm)}  - \nm \sum_{x \in \mathcal{X}} P_\nm(x) D(W_\theta (\cdot | x) \| P_\theta W_\theta) \right) \right. \nonumber \\
& ~~~~~~~~~~~~\left. \le  S - 2\gamma -   \frac{\log (N_\nm + 1)}{\sqrt{\nm}} \Big| X^\nm = \vect{x}_\nm \right\} =: f_{\theta, \nm}. \label{eq:new_LB7}
\end{align}
Since $V_{\theta, P_\theta}(\vect{x}_\nm) < + \infty$ and the third moment of $\frac{1}{n} \log \frac{W_\theta^\nm(Y_\theta^\nm|\vect{x}_\nm) }{(P_\theta W_\theta)^{\times \nm}(Y_\theta^\nm)}$ is also bounded (cf. \cite[Remark 3.1.1]{Han2003}, \cite[Lemma 62]{PPV2010}, \cite[Lemma 7]{Tomamichel-Tan2013}),  by the Berry-Ess\'een theorem and the {relations} in \eqref{eq:independent_sum_mean2} {and  \eqref{eq:independent_sum_variance2}}, we have
\begin{align}
\left| f_{\theta, \nm} - G\left(\frac{S-2\gamma -  \frac{\log (N_\nm + 1)}{\sqrt{\nm}} }{\sqrt{V_{\theta, P_\theta}(\vect{x}_\nm)}}  \right) \right| \le \frac{\nu_0}{\sqrt{\nm}} ~~(\forall \nm = n_1, n_2,\cdots),
\end{align}
where $G(\cdot)$ is defined as in \eqref{eq:Gaussian_dist} and $\nu_0 > 0$ is a positive {constant}. 
{Notice here that $V_{\theta, P_\theta}(\vect{x}_m) \rightarrow V_{\theta, P_\theta}$ as $m \rightarrow \infty$ owing to \eqref{eq:accumulation_dist}.}
{For $\theta \in \Theta_2$ with $V_{\theta, P_\theta} > 0$, we have $V_{\theta, P_\theta}(\vect{x}_m) > 0$ for all $\nm \ge \nm_1$ with some $\nm_1 > 0$.} 
Since $\log (N_\nm + 1) \le |\mathcal{X}| \log (\nm+1)$ and $G(\cdot)$ is continuous, {by letting $m \rightarrow \infty$} we obtain
\begin{align}
 \liminf_{\nm \rightarrow \infty} f_{\theta, \nm} &\ge  G\left(\frac{S-3\gamma }{\sqrt{V_{\theta, P_\theta}}}  \right)  \nonumber \\
 & = \Psi_{\theta, P_\theta} (S-3\gamma), \label{eq:A_LB2b}
\end{align}
where we have used the relation in \eqref{eq:Gaussian_dist} for the equality.
 For $\theta \in \Theta_2$ with $V_{\theta, P_\theta} = 0$, $G(z/\!\sqrt{V_{\theta, P_\theta}})$ is the step function which takes zero for $z < 0$ and one otherwise. 
Then, we have \eqref{eq:A_LB2b} for such $\theta \in \Theta_2$, too. 
Putting \eqref{eq:LB_alpha}, \eqref{eq:func_A}, \eqref{eq:new_LB7}, and  \eqref{eq:A_LB2b} together, we obtain 
\begin{align}
\liminf_{\nm \rightarrow \infty}  A^{(2)}_m & \ge \liminf_{\nm \rightarrow \infty} \int_{\Theta_2} dw(\theta) \alpha_{\theta, \nm} \nonumber \\
& \ge \liminf_{\nm \rightarrow \infty} \int_{\Theta_2} dw(\theta) f_{\theta, \nm} \nonumber \\
& \ge  \int_{\Theta_2} dw(\theta) \liminf_{\nm \rightarrow \infty} f_{\theta, \nm} \nonumber \\
& \ge  \int_{\Theta_2} {\Psi_{\theta, P_\theta} (S-3\gamma) \, dw(\theta)}  \nonumber \\
& \ge \inf_{P \in\Pi_{\bartheta,\Gamma}} \int_{\Theta_2} {\Psi_{\theta, P} (S - 3 \gamma) \, dw(\theta)} , 
\label{eq:beta_expansion2}
\end{align}
where we have used Fatou's lemma in the third inequality and the relation $P_\theta \in \Pi_{\theta, \Gamma} = \Pi_{\bartheta, \Gamma}$ in the last inequality. 
\end{itemize}

To finalize the evaluation of $A^{(2)}_m$ for both the two cases,  combining \eqref{eq:beta_LB2} and \eqref{eq:beta_expansion2} leads to
\begin{align}
\liminf_{\nm \rightarrow \infty}  A^{(2)}_m  & \ge \inf_{P \in\Pi_{\bartheta,\Gamma}} \int_{\{\theta \,|\, {c_{\theta, \Gamma}} = R\}} {\Psi_{\theta, P} (S - 3 \gamma) \, dw(\theta)}   \nonumber \\
& = \inf_{P \in\Pi_{\bartheta,\Gamma}} \int_{\{\theta \,|\, I(P, W_\theta) = R\}} {\Psi_{\theta, P} (S - 3 \gamma) \, dw(\theta)} \label{eq:beta_expansion4}
\end{align}
because $\Theta_2 = \{\theta \,|\, {c_{\theta, \Gamma}} = R\} = \{\theta \,|\, I(P, W_\theta) = R\}$ for any $P \in \Pi_{\bartheta, \Gamma}$ with $R = {c_{\bartheta, \Gamma}}$.

Now, we are in a position to synthesize all evaluations.
By the definition of achievability, it follows from \eqref{eq:new_LB3}, which means $\varepsilon_\nm \ge A^{(1)}_\nm + A^{(2)}_\nm -e^{-\sqrt{\nm} \gamma}$, \eqref{eq:beta_expansion1a}, and \eqref{eq:beta_expansion4} that
\begin{align}
\varepsilon &\ge \limsup_{n \rightarrow \infty} \varepsilon_n  \nonumber \\
 & \ge \limsup_{\nm \rightarrow \infty} \varepsilon_\nm \nonumber \\
  & \ge \limsup_{\nm \rightarrow \infty}  \left( A^{(1)}_\nm + A^{(2)}_\nm \right) \nonumber \\
 & \ge \liminf_{\nm \rightarrow \infty}  A^{(1)}_\nm + \liminf_{\nm \rightarrow \infty}  A^{(2)}_\nm \nonumber \\
 & \ge \int_{\{ \theta \,  | \, {c_{\theta, \Gamma}} < R \}} dw(\theta)  +\inf_{P \in\Pi_{\bartheta,\Gamma}} \int_{\{\theta \,|\, I(P, W_\theta) = R\}} {\Psi_{\theta, P} (S - 3 \gamma)  \,dw(\theta)}  \nonumber \\
 & = \inf_{P \in \Pi_{\bartheta, \Gamma}} \left\{ \int_{\{ \theta \,  | \, {c_{\theta, \Gamma}} < R \}} \, dw(\theta)  + \int_{\{\theta \,|\, I(P, W_\theta) = R\}} {\Psi_{\theta, P} (S - 3 \gamma) \, dw(\theta)} \right\}. \label{eq:WO_second_LB}
\end{align}
We note that $\Theta_1 = \{ \theta \,  | \, {c_{\theta, \Gamma}} < R \} = \{ \theta \,  | \, I(P, W_\theta) < R \}$ for any $P \in \Pi_{\bartheta,\Gamma}$ with $R = {c_{\bartheta, \Gamma}}$ due to the definition of $\Gamma$-well-orderedness, so it follows from \eqref{eq:WO_second_LB} that
\begin{align}
S - 3 \gamma & \le\sup_{ P \in \Pi_{\bartheta, \Gamma} } \sup \left\{ S \, \Big| \, \int_{\{\theta | I(P, W_\theta) < R \}} d w(\theta) +  \int_{\{\theta | I(P, W_\theta) =  R \}} {\Psi_{\theta, P}(S) \, d w(\theta)} \le  \varepsilon  \right\}. \label{eq:WO_second_2LB}
\end{align}
Since $\gamma > 0$ is arbitrary, we completed the proof of the converse part.
\QED

\section{{Concluding Remarks}} \label{sect:conclusion}

In this paper, we have established {the} coding theorem for the $(\varepsilon|\Gamma)$-capacity of mixed memoryless channels with general mixture.
For {mixed memoryless channels with general mixture}, a direct part of the second-order coding theorem has also been provided.
The class of $\Gamma$-well-ordered mixed memoryless channels, whose component channels are ordered according to their capacity with cost constraint $\Gamma$, {has been} introduced to {further} analyze the second-order  $(\varepsilon, R|\Gamma)$-capacity.
The $\Gamma$-well-orderedness allows us to establish a second-order converse theorem, which coincides with the direct theorem for {mixed memoryless channels with general mixture}.
The obtained results include several known results as special cases such as capacity characterizations for {mixed memoryless channels with general mixture} \cite{Ahlswede68,Han2003} and for regular decomposable channels with stationary memoryless components \cite{Winkelbauer71a}, an $\varepsilon$-capacity characterization for mixed memoryless channels with countable mixture \cite{Yagi-Nomura2014a}, and {second-order $(\varepsilon, R)$-capacity characterizations} for additive-noise channels with finite mixture \cite{PPV2011a} and for well-ordered memoryless channels with finite mixture \cite{Yagi-Nomura2014b}.

Tomamichel and Tan \cite{Tomamichel-Tan2014} {have} recently discussed mixed memoryless channels with finite $\Theta$ by treating them as memoryless channels with finite states.
In other words, channel state $\theta \in \Theta$ is selected with probability $w(\theta)$ before the transmission of a codeword of length $n$. 
In the scenario where the encoder and decoder can observe channel state $\theta$, characterizations for the $(\varepsilon|\Gamma)$-capacity and $(\varepsilon, R|\Gamma)$-capacity have been discussed.
Indeed, when $\Theta$ is finite and the encoder and decoder can access to the channel state information, the $(\varepsilon|\Gamma)$-capacity and $(\varepsilon, R|\Gamma)$-capacity are characterized as the natural counterparts of those in \eqref{eq:e-Cap_WO_general_mixture} and \eqref{eq:WO_2nd_order_general_mixture}, respectively, even for mixed memoryless channels whose component channels are not necessarily  $\Gamma$-well-ordered.
We can easily extend this result to mixed memoryless channels with general mixture (general states).

As noted in Sect.\ \ref{sect:proof_WO_mixed_DMC_2nd}, Lemma \ref{lem:mixed_lower_bound2} holds for mixed channels with general input and output alphabets ($\mathcal{X}$ and $\mathcal{Y}$), and we can also establish the converse part of the first-order coding theorem which corresponds to Theorem \ref{theo:general_mixed_DMC_e-cap} in the case with finite $\mathcal{X}$ and general $\mathcal{Y}$.
However, the proof of a direct part in this case may be trickier because we rely on the upper-decomposition technique of Lemma \ref{lem:mixed_upper_decomposition} (that is,  the method of types). 
Extensions of the established formulas for mixed channels with general input {and/or} output alphabets are interesting and practically important research subjects.


\appendices

\section{Proof of Lemma \ref{lem:dominant_component}} \label{append:proof_lem_dominant_component}

Given an arbitrary i.i.d.\ product probability distribution ${Q_\theta^n} $ on $\mathcal{Y}^n$, let ${Q^n}$ be
given as in \eqref{eq:output_dist_set}.
Since ${Q^n}(\vect{y})$ is the expectation of ${Q_\theta^n}(\vect{y})$ with respect to $w(\theta)$, Markov's inequality implies {that}
\begin{align}
\Pr \left\{ \theta \in {\Theta}(\vect{y}) \right\} \ge 1 - e^{-\sqrt[4]{n}} ~~(\forall \vect{y} \in \mathcal{Y}^n ).
\end{align}
We also have
\begin{align}
\Pr \left\{ \theta \in {\Theta}_n^c  \right\} &=  \Pr \left\{ \theta \in \cup_{k} {\Theta}(S_k)^c  \right\} \nonumber \\
&\le \sum_{k} \Pr \left\{ \theta \in {\Theta}(S_k)^c  \right\}  \le  (n+1)^{|\mathcal{Y}|}e^{-\sqrt[4]{n}}. \label{ineq_Phi0}
\end{align}
Here, $A^c$ denotes the complement of a set $A$.
Therefore, 
\begin{align}
\Pr \left\{ \theta \in {\Theta}_n  \right\} \ge 1 - (n+1)^{|\mathcal{Y}|}e^{-\sqrt[4]{n}}. \label{ineq_Phi}
\end{align}
{In a similar way,} we also have
\begin{align}
\Pr \left\{ \theta \in \tilde{{\Theta}}_n  \right\} \ge 1 - (n+1)^{|\mathcal{X}| \cdot |\mathcal{Y}|}e^{-\sqrt[4]{n}}.
\end{align}
{Then, it holds for ${\Theta}_n^* = {\Theta}_n \cap \tilde{{\Theta}}_n$ that} 
\begin{align}
\Pr \left\{ \theta \in {\Theta}_n^*  \right\} \ge 1 - 2 (n+1)^{|\mathcal{X}| \cdot |\mathcal{Y}|}e^{-\sqrt[4]{n}},
\end{align}
{thus,} yielding \eqref{eq:dominant_subset}.
\QED

\section{Proof of Lemma \ref{lem:mixed_upper_decomposition}} \label{append:proof_U_decomposition_lemma}
The proof is implicitly contained in Han \cite{Han2003}.
We summarize it here for {the} reader's convenience.

{For a given} $\gamma> 0$, we define a set
\begin{eqnarray}
D_n & = & \left\{ {\bf y} \in {\cal Y}^n \left| \frac{1}{n}\log P_{Y_\theta^n}({\bf y}) - \frac{1}{n} \log P_{Y^n}({\bf y})  \le - \frac{\gamma}{\sqrt{n}} \right.   \right\}
\end{eqnarray}
for $\theta \in \Theta$.
Then, it holds that
\begin{eqnarray}
\Pr\left\{ Y^n_\theta \in D_n \right\}  & = &  \sum_{{\bf y} \in D_n} P_{Y_\theta^n}({\bf y}) \nonumber \\
& \le & \sum_{{\bf y} \in D_n} P_{Y^n}({\bf y})  e^{-\sqrt{n}\gamma} \nonumber  \\
& \le & e^{-\sqrt{n}\gamma}.
\end{eqnarray}
%
Hence, for any {real number $z_n$} we have 
\begin{eqnarray} \label{eq:u-1}
\Pr\left\{  {-} \frac{1}{n} \log  {P_{Y^n}(Y^n_\theta)} \le z_n \right\}  
& \le & {\Pr\left\{ - \frac{1}{n} \log  {P_{Y^n}(Y^n_\theta)} \le z_n , \, Y_\theta^n \not\in D_n  \right\} +  \Pr\left\{ Y_\theta^n \in D_n \right\}} \nonumber  \\
& \le & \Pr\left\{ - \frac{1}{n} \log P_{Y^n_\theta}(Y_\theta^n) \le z_n + \frac{\gamma}{\sqrt{n}}   \right\} +  e^{-\sqrt{n}\gamma}
\end{eqnarray}
{for all $\theta \in \Theta$.}
By using the above inequality we have
\begin{eqnarray}
\Pr\left\{ \frac{1}{n} \log  \frac{W^n(Y_\theta^n|X^n)}{P_{Y^n}(Y^n_\theta)} \le z_n \right\}  
& = & \Pr\left\{ \frac{1}{n} \log {W^n(Y_\theta^n|X^n)} - \frac{1}{n}\log {P_{Y^n}(Y_\theta^n)} \le z_n   \right\} \nonumber  \\
& \le & \Pr\left\{ \frac{1}{n} \log {W^n(Y_\theta^n|X^n)} - \frac{1}{n}\log {P_{Y^n_\theta}(Y_\theta^n)} \le z_n + \frac{\gamma}{\sqrt{n}}   \right\} +  e^{-\sqrt{n}\gamma} \nonumber \\
& \le & \Pr\left\{ \frac{1}{n} \log {{W_\theta^n}(Y_\theta^n|X^n)} - \frac{1}{n}\log {P_{Y^n_\theta}(Y_\theta^n)} \le z_n + \frac{\gamma}{\sqrt{n}} + \frac{1}{{\sqrt[4]{n^3}}}  \right\} +  e^{-\sqrt{n}\gamma} \nonumber \\
& = & \Pr\left\{ \frac{1}{n} \log  \frac{W_\theta^n(Y_\theta^n|X^n)}{P_{Y_\theta^n}(Y^n_\theta)} \le z_n  + \frac{\gamma}{\sqrt{n}} + \frac{1}{{\sqrt[4]{n^3}}}\right\} + e^{-\sqrt{n}\gamma}
\end{eqnarray}
for $\theta \in \Theta^\ast_n$, where the last inequality is due to the inequality $W^n_\theta ({\bf y}|{\bf x}) \le e^{\sqrt[4]{n}}W^n({\bf y}|{\bf x})$ for $\theta \in \Theta^\ast_n$.
This completes the proof.
\QED

\section{Proof of Lemma \ref{lem:mixed_lower_decomposition}} \label{append:proof_L_decomposition_lemma}
This proof is also implicitly contained in Han \cite{Han2003} {in the case of $Q^n = P_{Y^n}$, where $P_{Y^n}$ denotes the output distribution on $\mathcal{Y}^n$ due to input $X^n$ via channel $W^n$}.
Similarly to (\ref{eq:u-1}), we obtain
\begin{eqnarray} \label{eq:u-2-1}
\Pr\left\{  \frac{1}{n}\log W^n(Y^n_\theta| X^n ) \le z_n \right\} \ge \Pr\left\{  \frac{1}{n}\log W_\theta^n(Y^n_\theta| X^n ) \le z_n - \frac{\gamma}{\sqrt{n}}\right\} - e^{-\sqrt{n}\gamma}
\end{eqnarray}
 for $\theta \in \Theta$.
Using {this inequality}, we have
\begin{eqnarray}
\Pr\left\{ \frac{1}{n} \log  \frac{W^n(Y_\theta^n|X^n)}{{Q^n}(Y^n_\theta)} \le z_n \right\}  
& = & \Pr\left\{ \frac{1}{n} \log {W^n(Y_\theta^n|X^n)} - \frac{1}{n}\log {{Q^n}(Y_\theta^n)} \le z_n   \right\} \nonumber  \\
& \ge & \Pr\left\{ \frac{1}{n} \log {W_\theta^n(Y_\theta^n|X^n)} - \frac{1}{n}\log {{Q^n}(Y_\theta^n)} \le z_n - \frac{\gamma}{\sqrt{n}}   \right\} -  e^{-\sqrt{n}\gamma} \nonumber \\
& \ge & \Pr\left\{ \frac{1}{n} \log {W_\theta^n(Y_\theta^n|X^n)} - \frac{1}{n}\log {{Q_\theta^n}(Y_\theta^n)} \le z_n - \frac{\gamma}{\sqrt{n}} - \frac{1}{{\sqrt[4]{n^3}}}  \right\} -  e^{-\sqrt{n}\gamma} \nonumber \\
& = & \Pr\left\{ \frac{1}{n} \log  \frac{W_\theta^n(Y_\theta^n|X^n)}{{Q_\theta^n}(Y^n_\theta)} \le z_n  - \frac{\gamma}{\sqrt{n}} - \frac{1}{{\sqrt[4]{n^3}}}\right\} -  e^{-\sqrt{n}\gamma}
\end{eqnarray}
for $\theta \in \Theta^\ast_n$, {where the last inequality is due to the inequality $Q_\theta^n ({\bf y}) \le e^{\sqrt[4]{n}}Q^n({\bf y})$ for $\theta \in \Theta^\ast_n$.}
Thus, we complete the proof.
\QED
\section{Proof of Lemma \ref{lem:mixed_lower_bound2}} \label{append:proof_of_new_lower_bound}

{For any given $(n, M_n, \varepsilon_n)$ code $\Code = \{ \vect{u}_1, \vect{u}_2, \ldots, \vect{u}_{M_n}\}$}, it follows from \eqref{eq:mixed_channel_general_measure} and \eqref{eq:ave_prob} that
\begin{align}
\varepsilon_n &=  \frac{1}{M_n} \sum_{i=1}^{M_n} W^n(D_i^c | \vect{u}_i)  \nonumber \\
 & =  \frac{1}{M_n} \sum_{i=1}^{M_n}  \int_{\Theta} dw(\theta) W_\theta^n(D_i^c | \vect{u}_i)   \nonumber \\
 & = \int_{\Theta}  dw(\theta) \left\{ \frac{1}{M_n} \sum_{i=1}^{M_n}  W_\theta^n(D_i^c | \vect{u}_i)   \right\}, \label{eq:mixed_LB} 
\end{align} 
where the equality in \eqref{eq:mixed_LB} is obtained by exchanging the integral and the sum of finitely many terms. 
Here, the term inside the brace $\{\cdot\}$ in \eqref{eq:mixed_LB} corresponds to the average error probability  {with the decoding region} $\{D_i \}_{i = 1}^{M_n}$ over $W_\theta^n$.
Then, a simple but key observation is that each of such terms indexed by $\theta \in \Theta$, which is characterized by the common decoding region $\{ D_i \}_{i =1}^{M_n}$,  may be lower bounded separately using another set depending on $\theta \in \Theta$.

Define the set
\begin{align}
B_{\theta, i} := \left\{ \vect{y} \in \mathcal{Y}^n \, \Big| \,  \frac{1}{n} \log \frac{W_\theta^n(\vect{y}|\vect{u}_i)}{{Q_\theta^n}(\vect{y})} \le  \frac{1}{n} \log M_n - {\eta} \right\}.
\end{align}
Then the term inside the brace $\{\cdot\}$ in \eqref{eq:mixed_LB} can be bounded as
\begin{align}
 \frac{1}{M_n} \sum_{i=1}^{M_n}  W_\theta^n(D_i^c | \vect{u}_i)  &\ge  \frac{1}{M_n} \sum_{i=1}^{M_n}  W_\theta^n(D_i^c \cap B_{\theta,i} | \vect{u}_i)  \nonumber \\
&   = \frac{1}{M_n} \sum_{i=1}^{M_n}  W_\theta^n(B_{\theta,i} | \vect{u}_i)  - \frac{1}{M_n} \sum_{i=1}^{M_n}  W_\theta^n(D_i \cap B_{\theta,i} | \vect{u}_i),   \label{eq:mixed_LB2}
\end{align} 
where the equality in \eqref{eq:mixed_LB2} follows from the relation
\begin{align}
D_i^c \cap B_{\theta,i} = B_{\theta,i}\setminus (D_i \cap B_{\theta,i}).
\end{align}
We focus on the second term in  \eqref{eq:mixed_LB2}.
By definition, every $\vect{y} \in B_{\theta, i}$ satisfies
\begin{align}
\frac{1}{M_n} W_\theta^n (\vect{y} | \vect{u}_i) \le {Q_\theta^n}(\vect{y}) e^{-n {\eta}}. \label{eq:set_B} 
\end{align}
Then the second term in \eqref{eq:mixed_LB2} is bounded as
\begin{align}
 \frac{1}{M_n} \sum_{i=1}^{M_n}  W_\theta^n(D_i \cap B_{\theta,i} | \vect{u}_i)
& =  \frac{1}{M_n} \sum_{i=1}^{M_n} \sum_{\vect{y} \in D_i \cap B_{\theta,i}} W_\theta^n(\vect{y} | \vect{u}_i) \nonumber\\
&   \le e^{-n {\eta}} \sum_{i=1}^{M_n}  \sum_{\vect{y} \in D_i \cap B_{\theta,i}}  {Q_\theta^n}(\vect{y}),   \label{eq:mixed_LB3} \\
&   \le e^{-n {\eta}} \sum_{i=1}^{M_n}   {Q_\theta^n}(D_i) =   e^{-n {\eta}} , \label{eq:mixed_LB4}
\end{align} 
where \eqref{eq:set_B} is used to obtain \eqref{eq:mixed_LB3}, and \eqref{eq:decoding_region} is used to obtain the equality in \eqref{eq:mixed_LB4}.

Plugging \eqref{eq:mixed_LB4} into \eqref{eq:mixed_LB2} yields\footnote{Inequality \eqref{eq:mixed_LB5} is {Hayashi-Nagaoka's} lower bound on the probability of decoding error, which has been originally established for the quantum channel setting \cite{Hayashi-Nagaoka2003}, for the component channel $W_\theta^n$. The derivation is essentially the same but slightly more direct than the original derivation (cf. \cite[Sect.\ IX-B]{Hayashi2009}).}
\begin{align}
 \frac{1}{M_n} \sum_{i=1}^{M_n}  W_\theta^n(D_i^c | \vect{u}_i)  &\ge  \frac{1}{M_n} \sum_{i=1}^{M_n}  W_\theta^n(B_{\theta,i} | \vect{u}_i)  -e^{-n {\eta}}.
  \label{eq:mixed_LB5}
\end{align} 
{Thus}, the left-hand side of \eqref{eq:mixed_LB} is lower bounded as
\begin{align}
\varepsilon_n &\ge \int_{\Theta} dw(\theta) \left\{ \frac{1}{M_n} \sum_{i=1}^{M_n} W_\theta^n(B_{\theta,i} | \vect{u}_i)   \right\}   -e^{-n {\eta}}, \label{eq:mixed_LB6} 
\end{align} 
which is equivalent to \eqref{eq:new_LB1}.
\QED

\section*{Acknowledgment}

The authors would like to thank Associate Editor, Prof.\ Aaron Wagner and the anonymous reviewers for their valuable comments which have improved the content of this paper.
This research was supported in part by MEXT under Grant-in-Aid for Scientific Research (C) No. 25420357 and No. 26420371.

\end{document}